\documentclass[prx,reprint,amsmath,amssymb,aps,superscriptaddress,showpacs]{revtex4-2}
\usepackage{blindtext}
\usepackage[colorlinks=true,allcolors=blue!92!black]{hyperref}
\usepackage{amsmath,amssymb}
\usepackage{graphicx}
\usepackage{amsfonts}
\usepackage{times}
\usepackage{color,xcolor}
\usepackage{bbm}
\usepackage{bm}
\usepackage{tabularx}
\usepackage{multirow}
\usepackage{mathtools}
\usepackage{xfrac}
\usepackage{latexsym}
\usepackage{mathrsfs}
\usepackage{slashed}
\usepackage{empheq}
\usepackage{dcolumn}
\usepackage{ulem}
\usepackage{cancel}
\usepackage{braket}
\usepackage[title]{appendix}
\usepackage{orcidlink}
\usepackage{dblfnote}
\usepackage{cancel}

\def\beq{\begin{equation}}
	\def\eeq{\end{equation}}
\def\bea{\begin{eqnarray}}
	\def\eea{\end{eqnarray}}
\newcommand{\q}{\mathbf{q}}
\newcommand{\p}{\mathbf{p}}
\newcommand{\x}{\mathbf{x}}
\def\k{\textbf{k}}

\def\inte{\textrm{int}}
\newcommand{\K}{\mathbf{K}}

\begin{document}
	
	\title{Gravitationally mediated entanglement of fermionic qubits: from static to dynamical limits}
	
	\author{M. Zarei\,\orcidlink{0000-0001-7744-2817}}
\email{Author to whom any correspondence should be addressed: {m.zarei@iut.ac.ir}  }
	\affiliation{Department of Physics, Isfahan University of Technology, 84156-83111 Isfahan, Iran}
	\affiliation{Quantum Technology Research Group, Isfahan University of Technology, Isfahan 84156-83111, Iran}
	\affiliation{Dipartimento di Fisica e Astronomia ``Galileo Galilei'' Universit\`a di Padova, I-35131 Padova, Italy}
	
	\author{M. Abdi\,\orcidlink{0000-0002-9681-5751}}
	\email{mehabdi@gmail.com}
	\affiliation{Department of Physics, Isfahan University of Technology, 84156-83111 Isfahan, Iran}
	\affiliation{Wilczek Quantum Center, School of Physics and Astronomy, Shanghai Jiao Tong University, Shanghai 200240, China}
	
	\author{N. Bartolo\,\orcidlink{0000-0001-8584-6037}}
	\email{nicola.bartolo@pd.infn.it}
	\affiliation{Dipartimento di Fisica e Astronomia ``Galileo Galilei'' Universit\`a di Padova, I-35131 Padova, Italy}
	\affiliation{INFN, Sezione di Padova, I-35131 Padova, Italy}
	\affiliation{INAF - Osservatorio Astronomico di Padova, I-35122 Padova, Italy}
	
	\author{S. Matarrese\,\orcidlink{0000-0002-2573-1243}}
	\email{sabino.matarrese@pd.infn.it}
	\affiliation{Dipartimento di Fisica e Astronomia ``Galileo Galilei'' Universit\`a di Padova, I-35131 Padova, Italy}
	\affiliation{INFN, Sezione di Padova, I-35131 Padova, Italy}
	\affiliation{INAF - Osservatorio Astronomico di Padova, I-35122 Padova, Italy}
	\affiliation{Gran Sasso Science Institute, I-67100 L'Aquila, Italy}
	
	\begin{abstract}
		We employ the quantum Boltzmann equation to analyze the gravitationally generated entanglement between two remote qubits by considering two explicit microscopic models. A graviton propagator is employed as the mediator of the interactions, while the qubits are considered in a spatial superposition state. Such a setup, in the case of any entanglement generation, could potentially offer experimental evidence for the quantization of gravity. By treating the qubits as spin-1/2 particles in wave packets, we establish that the entanglement arises from forward scattering processes involving graviton exchanges. In our study, we consider both static and dynamical limits of the propagator and show that only in the dynamical limit such entangled states can be generated. 
	These effects are observed to be diminished as the wave packet size increases. 
		Our findings sheds more light into the gravity mediated entanglement between two spin-1/2 particles. 
	\end{abstract}
	
	\maketitle

\section{Introduction} 
The question of whether gravity is fundamentally a quantum force remains one of the most profound challenges in modern physics.
 Recent theoretical and experimental proposals have revived this debate once again, particularly those investigating gravitationally induced entanglement between massive objects with embedded spins~\cite{Bose:2017nin,Marletto:2017kzi} (see~\cite{Bose:2023gwh, Marletto:2024ltk} for recent reviews and also appendix \eqref{appendix:A} and \eqref{appendix:B}).
In most of these works a Newtonian gravitational model is employed to investigate its quantum nature through LOCC (Local Operation and Classical Communication) theorem. This theorem implies the quantum state of a bipartite system remains separable if only local operations on the subsystems that might be coordinated via a classical communication channel are performed. In other words, such LOCC operations
are not sufficient to create an entangled state \cite{Nielsen}. 
A central point of discussion, however, is whether the observation of entanglement due solely to the Newtonian potential, which originates from a non-dynamical component of the gravitational field, is sufficient to establish the quantization of gravity~\cite{ Vicentini:2024noi, Abrahao:2023lle, Arani:2023ljw, Colas:2024xjy, Sugiyama:2024equ, Gallock-Yoshimura:2024wpz, Chevalier:2020uvv, Bassi:2017szd, Kent:2021ljj,bose2017spin,marletto2017gravitationally,Krisnanda:2019glc,Christodoulou2019,danielson2022gravitationally,galley2022no,christodoulou2023locally,Carney:2018ofe, kafri2013noise, Kafri:2014zsa, kafri2015bounds, Bahrami:2015wma, anastopoulos2015probing, bose2017spin, marletto2017gravitationally, haine2021searching, Qvarfort:2018uag, Carlesso:2019cuh, PhysRevA.101.063804, howl2021non, Matsumura:2020law, Pedernales:2021dja, liu2021gravitational, Datta:2021ywm, Trillo:2024kqz}. 
Specifically, the argument, still debated~\cite{Hall:2017nzl,martin2023gravity}, shows that if gravity is classical in the sense of being a LOCC then it cannot entangle two initially separated  systems~\cite{Horodecki:2009zz}; therefore if any entanglement is detected, gravity as the mediator must be of a quantum nature.

Nevertheless, there are arguments about the insufficiency of the Newtonian generated entanglement alone in establishing the quantization of the physically relevant, propagating degrees of freedom of the gravitational field~\cite{Anastopoulos:2020cdp, Anastopoulos:2018drh}.
In Ref.~\cite{Anastopoulos:2022vvb}, it is suggested that treating quantum field theory (QFT) as a quantum information channel oversimplifies its fundamental structure and neglects key physical principles. Unlike standard quantum information theory, which assumes separable systems and is largely non-relativistic, QFT inherently entangles the degrees of freedom of emitters, receivers, and mediating fields, necessitating the use of dressed states \cite{Gerry}. 
On the other hand, counterarguments propose that a consistent theoretical framework cannot simultaneously accommodate entanglement from Newtonian interaction while maintaining an unquantized metric~\cite{Belenchia2018, Carney:2021vvt,Martin-Martinez:2022uio,Reginatto:2018ksa,Chen:2024xvm}. 

%Despite the routine observation of quantum superpositions in microscopic particles like electrons and atoms, extending these phenomena to truly macroscopic scales remains challenging \cite{Anastopoulos:2020cdp,Ma2021,Bild2023}.
Despite the extreme experimental challenges in preparing macroscopic objects in a superposition state, observation of spatial quantum superpositions in microscopic particles like electrons and atoms, is routinely done in the labs~\cite{Anastopoulos:2020cdp,Ma2021,Bild2023}.
Therefore, a thorough understanding of the gravity-induced entanglement in microscopic systems, and thus, through valid microscopic models seems necessary and yet hitherto widely overlooked. This specifically applies to the fermionic systems.
In any case, the decoherence hinders the ability to detect such delicate effects.
%The heaviest objects demonstrated in quantum superposition so far are macromolecules with masses of the order of $10^{-23} \, \text{kg}$. Proposals to push this boundary towards larger masses of $10^{-14} \, \text{kg}$ using matter-wave interferometry or Stern-Gerlach setups face significant scalability issues. However, novel techniques leveraging diamagnetic repulsion show promise in extending spatial superposition sizes, though key challenges remain in initializing large superpositions and achieving sufficient velocity separation~\cite{Marshman:2018upe, Marshman:2021wyk, Zhou:2022frl, Pedernales_2020, Zhou:2022epb, Zhou:2022jug, Zhou:2024voj, Braccini:2024fey}.  A primary focus of spatial superposition states, in both experimental and theoretical contexts, is decoherence. 
%In the proposed experiment \cite{Bose:2017nin,Marletto:2017kzi}, decoherence hinders our ability to detect quantum gravity-induced entanglement. 
As a result, developing methods to mitigate these decoherence effects is of paramount importance~\cite{Rijavec:2020qxd, Sugiyama:2022ixw,Hidaka:2022gsv,Gunnink.Gravitational.2023,Wan_2016, vandeKamp:2020rqh, Toros:2020dbf, Wood:2021icq, Schut:2021svd, Tilly.Qudits.2021, Sharifian:2023jem}.
%Related studies have been carried out such as the effect of decoherence on the gravitationally induced entanglement \cite{Rijavec:2020qxd, Sugiyama:2022ixw,Hidaka:2022gsv,Gunnink.Gravitational.2023,Wan_2016, vandeKamp:2020rqh, Toros:2020dbf, Wood:2021icq} and methods for overcoming decoherence \cite{Schut:2021svd, Tilly.Qudits.2021}. 
%The decoherence of a spatially superposed \textit{spin degree of freedom} caused by spin-magnetic field coupling. 
%This is particularly interesting because the embedded spin in a mass is responsible for generating spatial superposition, yet it cannot avoid the decoherence from the vacuum fluctuations of a quantum electromagnetic field even if the mass is electrically neutral. 
%See \cite{Ford.electron.coherence.1993, Ford.electron.coherence.1997, Breuer.coherence.2001, Mazzitelli.Decoherence.2003} for the decoherence of an electron coupled to a quantum electromagnetic field, and \cite{Sharifian:2023jem} for gravity-induced decoherence on a spin-$\frac12$ qubit system.

In this work, we investigate entanglement generation via graviton exchange at the microscopic level between qubit systems composed of spin-1/2 particles, which is applicable to elementary particles as well as large atoms with definite spin states.
We investigate two microscopic models for the gravity-mediated interaction Hamiltonian between spatially distant particles and employ the quantum Boltzmann equation (QBE) to analyze the system dynamics through the forward scattering mechanism.
The forward-scattering term of the QBE has previously provided crucial insights into neutrino mixing~\cite{Mikheev:1987qk, Sigl:1993ctk}, cosmic microwave background polarization~\cite{Kosowsky:1994cy, Bavarsad:2009hm, Bartolo:2018igk, Bartolo:2019eac, Hoseinpour:2020hic}, and gravitational waves~\cite{Zarei:2021dpb}, with recent extensions to wavepackets~\cite{Sharifian:2025olk}.
We find that a gravity-mediated-entanglement between two massive spin-1/2 fermions is only possible when the dynamical part of the propagator is taken into account.
Furthermore, depending on the microscopic model in action, the dynamical gravity induced entanglement occurs in different particle mass regimes.
In the passive model where the spinors are naturally couple through the gravity mass of the particles determines the rate of entanglement generation.

\section{Interaction of two spin-1/2 systems via graviton}
\begin{figure}
	\centering
	\includegraphics[scale=0.26]{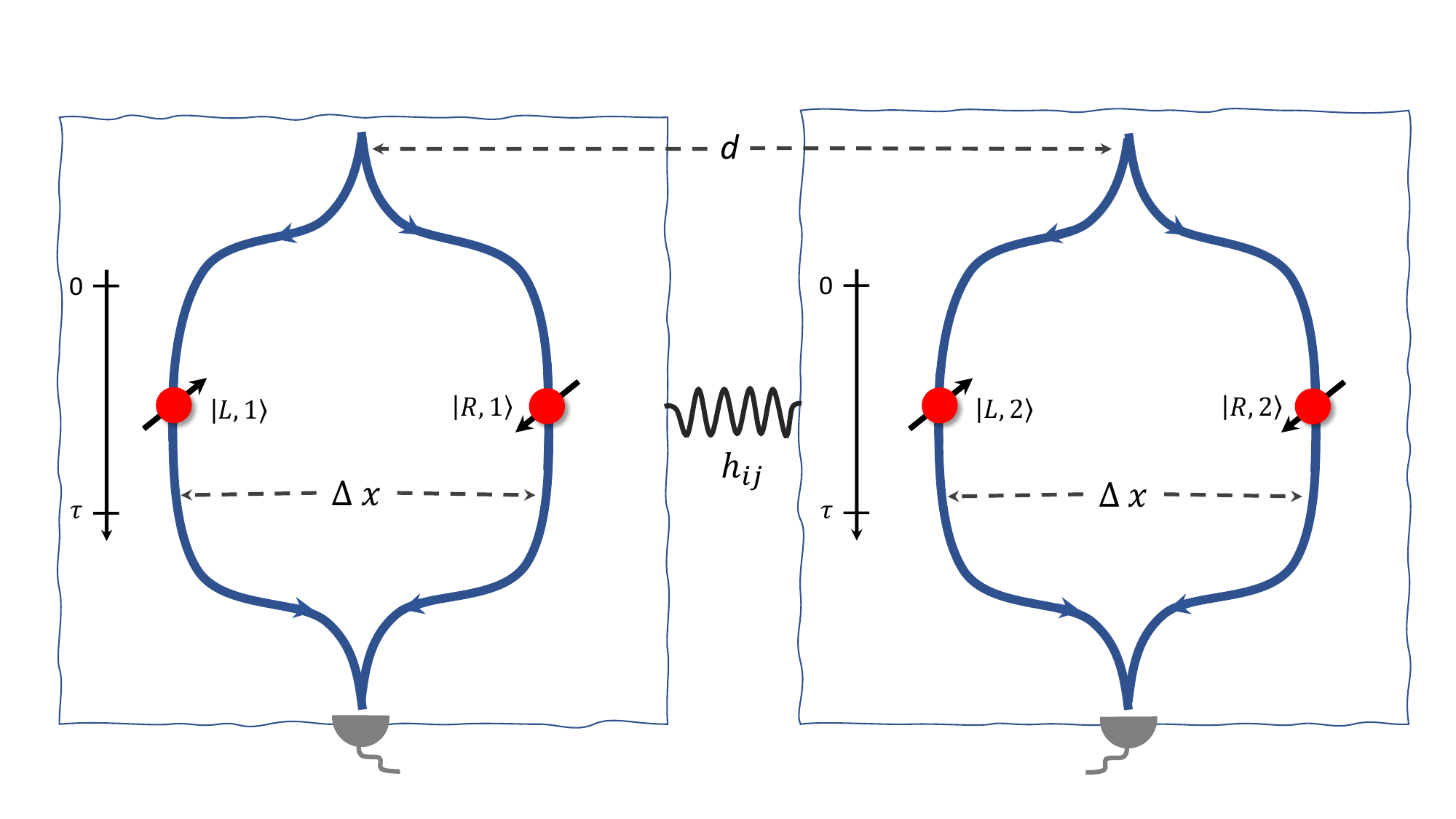} 
	\caption{Adjacent interferometers probe the quantum nature of gravity: Two test masses are placed adjacently in a spatial superposition of localized states 
		$|L\rangle$ and $|R\rangle$ through two Stern-Gerlach interferometers that split the motional states of the test masses in a spin-dependent manner, preparing states $|L,j\rangle+ |R,j\rangle$ for $j=1,2$. }
	\label{1}
\end{figure}

In the QBE framework, interactions are described using quantum fields and S-matrix elements~\cite{Kosowsky:1994cy, Bavarsad:2009hm, Bartolo:2018igk, Bartolo:2019eac, Hoseinpour:2020hic}. Here, we model qubit-qubit interactions by considering two spin-$\frac{1}{2}$ particles, described by a spinor field $\psi$, interacting via virtual graviton exchange. This interaction manifests as a scattering process between the two fermion particles (see Fig.~\ref{1}).
In this work we are interested in the low energy limit effects. Hence, in order to recover the qubit limit, the interaction is reduced to involve only spin operators in the non-relativistic regime.
Moreover, we parameterize the coupling of the graviton tensor field $h_{\mu\nu}$ to fermion matter fields in two different way.
That is, via the following interaction terms: $\kappa h_{\mu\nu} (\partial^\mu \bar{\psi}) \gamma^\nu \psi$ (Model I), and $\kappa h_{\mu\nu} A^\mu \bar{\psi} \gamma^\nu \psi$ (Model II), where $\kappa^2=16\pi G$, with $G$ being the gravitational constant, $\gamma^\mu$ the Dirac matrices, and $ A^\mu $ an external electromagnetic field.
While the former can be seen as a passive interaction that happens naturally, the latter symbolises an externally activated effect, which in this case requires an electromagnetic fields as the catalyst.
In the following, we present the Hamiltonian for Model I and analyze it. Given its negligibility, the Hamiltonian and detailed analysis for Model II is given in Appendix \ref{appendix:E}.

\subsection{Hamiltonian for Model I}
First, we consider the model in which two fermions interact via graviton exchange, whose dynamics is described by the following effective Hamiltonian
\begin{eqnarray}
	\hat{H}_{\textrm{int}}&=&\tfrac{1}{4}\kappa^2\int d^3x\, d^3x'\,  D^{\mu\nu\mu'\nu'}(\mathbf{x}-\mathbf{x}')
	\nonumber\\ &&\times 
	\partial_{\mu'}\bar{\psi}^-(\mathbf{x}')
	\gamma_{\nu'}  \psi^+(\mathbf{x}')
	\partial_{\mu}\bar{\psi}^-(\mathbf{x})
	\gamma_{\nu}  \psi^+(\mathbf{x})
	\nonumber\\ &&\times 
	VV'\delta_{\sigma_0}^3(\mathbf{x}-\bar{\mathbf{x}}_r)\delta_{\sigma'_0}^3(\mathbf{x}'-\bar{\mathbf{x}}'_s)~, \label{Hmodel1}
\end{eqnarray}
where $V$ and $V'$ represent spatial volumes within which the interaction is appreciable and $D^{\mu\nu\mu'\nu'}(\mathbf{x}-\mathbf{x}')$ is the graviton propagator.
Here, we treat matter states as \textit{wave packets} with finite spatial extensions $\sigma_0$ and $\sigma'_0$. To account for the localization, we introduce the following Gaussian distribution function centered at $\bar{\mathbf{x}}_r$ and with the width $\sigma_0$
\begin{eqnarray}\label{eq:delta4bar}
	\delta^3_{\sigma_0}(\mathbf{x}-\bar{\mathbf{x}}_r)=\frac{1}{(2\pi \sigma^2_0)^{3/2}}\exp\left[ - \frac{(\mathbf{x}-\bar{\mathbf{x}}_r)^2}{2\sigma_0^2} \right]~.
\end{eqnarray}
Similarly, $\delta^3_{\sigma'_0}(\mathbf{x} - \bar{\mathbf{x}}'_s)$ parametrizes the spatial extension of the second particle about $\bar{\mathbf{x}}'_s$.
Note that these expressions generally depend on the spin state of the particles~\cite{Sharifian:2025olk, Breuer2002}.
The spinor field $\psi$ can be decomposed into creation and annihilation components as  $\psi(x) = \psi^+(x) + \psi^-(x)$, where
\begin{equation}
	\psi^+(x) = \int d\mathbf{p} \sum_{r} u_r(t) \hat{c}_r(\mathbf{p}) e^{-i(p^0 t - \mathbf{p} \cdot \mathbf{x})}~,
\end{equation}
The annihilation component follows as $\psi^- = (\psi^+)^\dagger$.
The graviton propagator in the harmonic (de Donder) gauge in the momentum space is given by~\cite{Donoghue:1995cz}
\begin{align}
	D^{\mu\nu\mu'\nu'}= \frac{-1}{|\mathbf{K}|^2}(\eta^{\mu \mu' } \eta^{\nu \nu' } + \eta^{\mu \nu'} \eta^{\nu \mu' } -\eta^{\mu \nu } \eta^{ \mu' \nu' })~,
\end{align}
where $\eta^{\mu\nu}$ is the metric tensor and
 $\K$ is the exchange momentum. 
By considering the $(03,03)$ component of the propagator, we focus on a mixed space-time component of the gravitational field, $h_{03}$. This mode corresponds to a transverse-traceless gravitational perturbation, incorporating both spatial and temporal variations. In contrast, the $(00,00)$ component captures purely static effects.
Below we investigate effect of these two scenarios separately by refering to the former case as the spatially transverse limit, while the latter is the static limit.

\section{Quantum Boltzmann equation} 
The quantum Boltzmann equation (QBE) describes the evolution of a quantum system interacting with its environment~\cite{Kosowsky:1994cy,Bavarsad:2009hm,Bartolo:2018igk,Bartolo:2019eac,Hoseinpour:2020hic,Zarei:2021dpb}. Here, we extend this formalism to investigate entanglement dynamics between two quantum systems, $A$ and $B$, providing a generalized framework for their correlated evolution.
To track the evolution of entanglement, we apply a modified QBE to the $4\times 4$ density matrix $\rho_{IJ}$
\begin{align} 
	\left[(2\pi)^3\delta^3(0)\right]^2\dot{\rho}_{IJ}(\mathbf{k},t) &=i\left\langle\big[\hat{H}_{\textrm{int}}(0),\hat{\mathcal{D}}_{IJ}(\mathbf{k})\big]\right\rangle  \nonumber\\
	&\hspace{-2cm}-\int_{0}^{\tau}\hspace{-2mm}ds\left\langle\Big[\hat{H}_{\textrm{int}}(s),\big[\hat{H}_{\textrm{int}}(0),\hat{\mathcal{D}}_{IJ}(\mathbf{k})\big]\Big]\right\rangle~,\label{QBE1}
\end{align}
where the capital indices are $I,J=1,\dots,4$ and $\hat{\mathcal{D}}_{IJ}$ is the number operator.
In this formalism the expectation value of an operator $\hat{O}$ is given by
\begin{equation} \label{exp1} \langle \hat{O}(\mathbf{k}) \rangle = \mathrm{tr}[\rho \hat{O}(\mathbf{k})] =\int d \p\left\langle\mathbf{p}\middle| \hat{\rho} \hat{O}(\mathbf{k}) \middle| \mathbf{p} \right\rangle,
\end{equation}
with $d\mathbf{p} = d^3p/(2\pi)^3$. The density operator is defined as
\begin{equation}
\hat{\rho}=\int d\p\rho_{IJ}(\mathbf{p})\hat{\mathcal{D}}_{IJ}(\mathbf{p})~.
\end{equation}
In Eq.~\eqref{QBE1}, the first term on the right-hand side represents forward scattering, where the momenta remain unchanged but phase shifts alter the quantum coherence. The second term describes collision or decay processes, which induce decoherence. In this work, we put the focus on the forward scattering contribution to the entanglement dynamics.

\subsection{Static limit}
We now delve deeper into the components of QBE, focusing on the intricate details of the forward scattering term for both models I and II. We will explore how the forward scattering term primarily affects the phase of the quantum states, resulting in phase shifts that depend on the mass and spatial separation between the interacting systems. This analysis is crucial for understanding how gravitational interactions influence the quantum coherence of two interacting systems.
In the static limit, one sets $ \mu = \nu = \mu' = \nu' = 0 $ in the graviton propagator $ (D^{0000})  $ in the effective Hamiltonians \eqref{Hmodel1} and \eqref{Hmodel2}.
In Model I, we assume $ \partial_0 \psi \approx 0,$  
while in Model II, we set $ A_0 = 0 $ (details are presented in \eqref{appendix:E}). 
The main reason for this assumption is that the term $ \bar{\psi} \gamma^0 \psi $ sums to zero over all spin states in both models, and hence, the interaction Hamiltonian becomes vanishing.
Therefore, the QBE analysis leads to the conclusion that the density matrix elements remains unchanged over time
\begin{eqnarray} \label{QEPQBE1}
\dot\rho_{IJ}=0~.
\end{eqnarray}
This implies that the qubits undergo no joint evolution and since we assume that they are initialized in a separable state no entanglement is generated in this scenario.
In other words, classical gravity, particularly in the static limit, does not induce entanglement between quantum systems.
This is indeed consistent with the LOCC theorem, where entanglement cannot be generated by classical mediators.

\subsection{Spatially transverse limit}
We now consider the dynamical transverse-traceless case, focusing on the $D^{03 03}$ component of the graviton propagator. Here, the indices $0$ and $3$ correspond to the time and third spatial components, respectively. We assign the $z$-axis as the third spatial direction in our chosen geometrical configuration (see Fig.~\ref{1}).
In this case, the $h_{03}$ component describes a transverse-traceless mode of the gravitational perturbation. Unlike a purely static effect, this perturbation includes both spatial and temporal variations, affecting the interaction dynamics in a nontrivial manner.
To analyze this regime, we first calculate the forward scattering amplitude for Model I. The same procedure is followed for Model II.

%By invoking the S-matrix formalism, we express the effective interaction Hamiltonian in Fourier space from Eq.~\eqref{Hmodel1}, taking the transverse-traceless limit, as follows
%\begin{align}
%\hat{H}_{\inte}^{\rm (I)} &=- \frac{1}{4}\kappa^2 VV' \sum_{rr',ss'}
%\int d\mathbf{x} \, d\mathbf{x}' \, d\p \, d\p' \, d\q \, d\q' \, d\K \, \frac{p^0 q^0 }{|\K|^2} \, \notag \\
%&\quad \times
%\delta^3_{\sigma_{0}}(\x-\bar{\x}_r) \delta^3_{\sigma'_{0}}(\x'-\bar{\x}_s') e^{i(\K+\p-\p')\cdot \x} e^{i(\q-\q'-\K)\cdot \x'}  \notag \\
%&\quad \times  	 
%\bar{u}_{s'}(\q') (I\otimes \sigma^3) u_s(\q) \bar{u}_{r'}(\p') (I\otimes \sigma^3) u_r(\p)   \notag %\\
%&\quad \times
%b_{s'}^\dag(\q')b_{s}(\q)a_{r'}^\dag(\p')a_r(\p)~,
%\label{effH5}
%\end{align}
%where $I$ and $\sigma^3$ denote the identity and the third Pauli matrix, respectively. 
%We now proceed by substituting the interaction Hamiltonian \eqref{effH5} into the forward scattering term—the first-order term in the interaction Hamiltonian—of the quantum Boltzmann equation \eqref{QBE1}. This substitution enables us to determine the evolution of $\rho_{IJ}$ up to the second order in $\kappa$, see Appendix~C\ref{appendix:density} for a detailed derivation.
We now substitute the interaction Hamiltonian for Model I, $\hat{H}_{\inte}$, into the forward-scattering term of the QBE, the first line in Eq.~\eqref{QBE1}. After performing the integrations and the expectations values one arrives at the following equation that describes dynamics of the two-particle density matrix
\begin{equation} 
\dot{\rho}_{IJ} =- \frac{i \kappa^2}{16\pi} m_1m_2F(i,j,k,l)~,
\label{rhodynamics}
\end{equation} 
where we have introduced the $F$ function---with folded indices that follows the tensor product index mapping---as
\begin{align}
	F(i,j,k,l) &=
 \sum_{rr',ss'}  \frac{1}{R_{sr}}\bar{u}_{s'}(\k) (I \otimes \sigma^3) u_s(\k) \\
&\times \bar{u}_{r'}(\k) (I \otimes \sigma^3) u_r(\k)  [ \delta_{ri} \,\delta_{sk} \,  \rho^A_{r'j}(\k) \rho^B_{s'l}(\k) \nonumber\\
&\hspace{2cm}- \delta_{jr'} \, \delta_{ls'} \, \rho^A_{ir}(\k)\rho^B_{ks}(\k) ] \, \nonumber.  \label{ttdynamics}
\end{align}
Here, $I$ and $\sigma^3$ denote the identity and the third Pauli matrix, respectively.
$R_{sr} = |\bar{\mathbf{x}}_r - \bar{\mathbf{x}}'_s|$ is the distance between the corresponding spatial states, which in turn are dictated by the spin states.
Whereas $\rho^A$ and $\rho^B$ are the density matrices of the $A$ and $B$ subsystems.
The details of these calculations are provided in appendix \eqref{appendix:C}.
We now assume that the two particles are initially prepared in a superposition state through the Stern-Gerlach setup. 
In other words, $\rho_{IJ}(0)$ denotes the initial $4 \times 4$ density matrix with all entries equal to $1/4$. 
One solves the differential equation \eqref{rhodynamics} with this initial state to find the resulting entangled total density matrix as the following
\begin{widetext}
\begin{eqnarray} 
	\rho_{[IJ]}(\tau) = \frac{1}{4}
\begin{pmatrix}
	1 & e^{-i\Delta \phi_{LR}} & e^{-i\Delta \phi_{RL}} & 1 \\
	e^{i\Delta \phi_{LR}} & 1 & e^{-i(\Delta \phi_{RL} - \Delta \phi_{LR})} & e^{i\Delta \phi_{LR}}  \\
	e^{i\Delta \phi_{RL}} & e^{i(\Delta \phi_{RL} - \Delta \phi_{LR})} & 1 & e^{i\Delta \phi_{LR}} \\
	1&e^{-i\Delta \phi_{RL}}& e^{-i\Delta \phi_{LR}} & 1
\end{pmatrix}
~,
\end{eqnarray}
\end{widetext}
where $\Delta \phi_{RL}$ and $\Delta \phi_{LR}$ are given in Appendix~\ref{appendix:A}. 

The precise calculations presented in Appendix \ref{appendix:D} shows that the induced phase due to the graviton propagator is 
\begin{equation}
 \phi_{\text{G}} \approx \frac{Gm_1m_2\tau}{d\hbar } \text{erf} \left( \frac{d}{\sqrt{2} \sigma_0} \right)~,
\end{equation}
where for $d \gg \sigma_0$ one retains $\phi_{\text{G}} \approx \frac{G m_1 m_2 \tau}{\hbar d}$.
This result is consistent with the findings of Ref.~\cite{Bose:2017nin,Marletto:2017kzi}, which demonstrate the generation of entanglement between massive spinless particles in the static limit.
In our analysis, we construct a \( 4 \times 4 \) density matrix \( \rho_{IJ} \) by assuming an initially separable state which is given as the tensor product of two \( 2 \times 2 \) density matrices, \( \rho^A \) and \( \rho^B \), corresponding to two subsystems \( A \) and \( B \). The elements of \( \rho_{IJ} \) over time are then systematically determined by combining the elements of \( \rho^A \) and \( \rho^B \) and solving for the QBE.
This approach enables a direct characterization of the entanglement dynamics induced by graviton exchange.
The static limit supports the expectation that classical gravity should not generate a phase shift, reinforcing the notion that any observed quantum effects must originate from the dynamical properties of gravity. In contrast, the dynamical transverse limit introduces time-dependent corrections, leading to nontrivial phase evolution and potential revivals in quantum coherence. 

\begin{figure}[tb]
\includegraphics[width=\columnwidth]{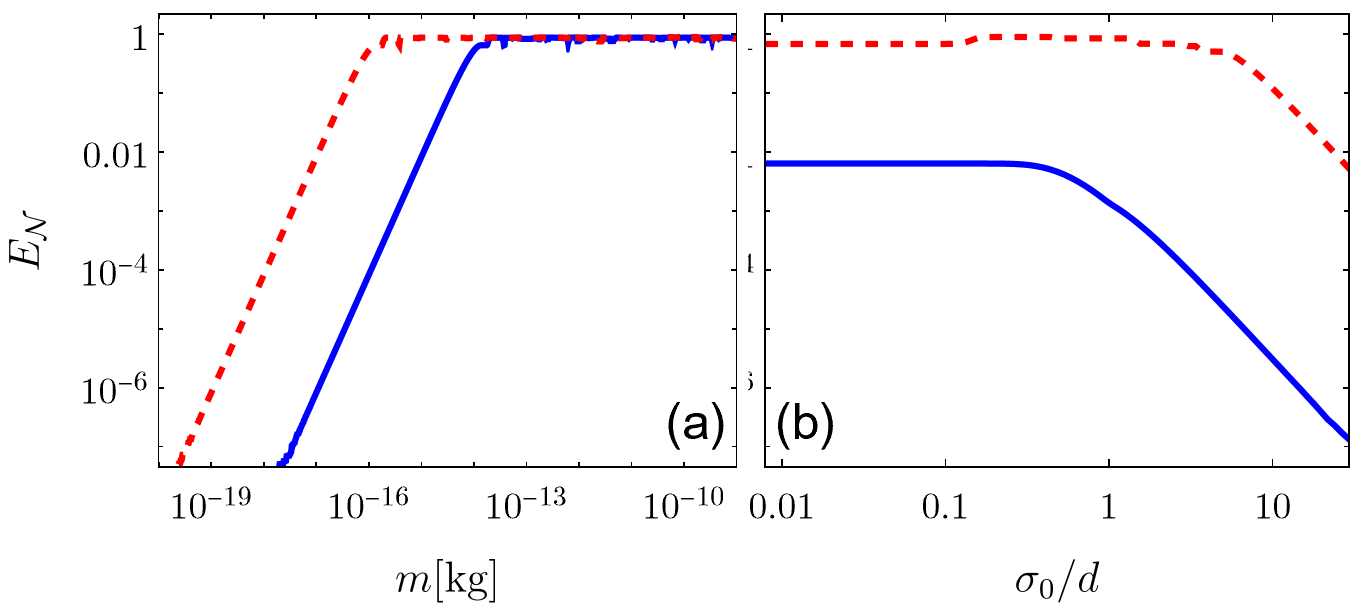}
\caption{Gravitationally induced entanglement: (a) The logarithmic negativity is shown as a measure of the entanglement built up between the particles as a function of their mass. Here, we have considered two set of parameters: $\{d,\tau\}=\{100~\text{nm}, 10^{-3}~\text{s}\}$ solid blue lines and $\{d,\tau\}=\{10~\text{nm}, 1~\text{s}\}$ for the red dashed lines. In both sets $\Delta x=d/2$ and $B_3=1$~T. (b) The logarithmic negativity is plotted as a function of the spatial distribution of the particle wave packets for the fixed mass of $m=10^{-15}$ kg.}\label{fig:models}
\end{figure}

In particular, here we consider two different setups suitable for such experiments.
One with $\{d,\tau\}=\{100~\text{nm}, 10^{-3}~\text{s}\}$ which is rather close to the state of the art experiments.
Meanwhile, we also consider the following parameter set that might be attainable in the future experiments: $\{d,\tau\}=\{10~\text{nm}, 1~\text{s}\}$.
In both cases we set the extend of the superposition to $\Delta x=d/2$, which means even in the closest branches the two particles remain apart from each other by $\tfrac{d}{2}$.
We compute the logarithmic negativity as a measure of entanglement which is defined as $E_\mathcal{N}=\log_2\Vert\rho^\Gamma\Vert_1$, where $\Vert \cdot \Vert_1$ is the trace norm and $\rho^\Gamma$ represents the partial transposition of the density matrix~\cite{Plenio2005}.
In Fig.~\ref{fig:models} the entanglement measure is plotted for the two above parameter sets as a function of mass and the spatial extent of the particles $\sigma_0$.
It is obvious from the plots that for Model~I a larger mass gives a higher entanglement.
We also investigate the effect of spatial diffusion of the particles and find that when the particles are localized in their paths a higher entanglement is attained, see Fig.~\ref{fig:models}(b).

\section{Conclusion}
We present a dynamical equation that describes the entanglement two spin-\(\frac{1}{2}\) systems, $A$ and $B$ to gravitational interaction mediated by a graviton propagator. Using an extended formulation of QBE, we analyzed the entanglement generation mechanism. Our results indicate that the entanglement effect reported in Refs. \cite{Bose:2017nin,Marletto:2017kzi} corresponds to forward scattering via graviton exchange in the dynamical regime, with an interaction strength proportional to the masses of both particles.  

The induced phase arising from gravitational interactions mediated by graviton exchange in the presence of a background magnetic field (Model II) is given in Eq.~\eqref{E3} and is subdominant compared to the pure gravitational interaction. Consequently, the presence of the magnetic field does not influence the entanglement induced by graviton exchange.   

In addition, we have examined the induced phase arising from interactions mediated by photon exchange. The details are presented in Appendix~\ref{appendix:F}. As discussed there, the spin--spin interaction leads to an induced phase $\phi_{\mathrm{SS}}$, which is dominant for the distances considered in this work and used in the analysis shown in Fig.~\ref{fig:models}, namely $d \sim 10^{-8}\,\mathrm{m}$ for particles with a mass around $m \sim 10^{-15}$ kg. The spin-spin interaction becomes negligible only for larger distances ($d \gtrsim 10^{-6}$ m) and or higher particle masses $m \gtrsim 10^{-13}\,\mathrm{kg}$.

Furthermore, the Casimir-Polder interaction constitutes as yet another potentially relevant background in such experiments. For the mass and separation scales considered here, the Casimir force can be comparable to or larger than the gravitational interaction if left unsuppressed. Nevertheless, it has been shown that the Casimir contribution as well as any other forces with electromagnetic root can be efficiently mitigated by an approperiate electromagnetic shielding, geometric symmetry, or differential measurement schemes, ensuring that it does not affect the entanglement generated by gravity.

Moreover, recent work by Aziz and Howl \cite{Aziz:2025ypo} has revisited the question of gravitationally induced entanglement by demonstrating that classical theories of gravity may, in principle, generate entanglement through physically local processes. Their results highlight that the mere observation of entanglement is not sufficient to unambiguously establish the quantum nature of gravity and emphasize the importance of understanding the microscopic origin and scaling of entanglement generation. In this context, our analysis provides complementary insight by elucidating the underlying mechanisms responsible for entanglement production and helping to clarify the conditions under which genuinely quantum gravitational effects can be isolated.

\begin{acknowledgements}
MZ thanks INFN and department of Physics and Astronomy “G. Galilei” at University of Padova for warm hospitality while this work was done.
\end{acknowledgements}

%\nocite{apsrev41Control} % for adding title of articles to the references
%\bibliographystyle{apsrev4-2}

\bibliography{refs.bib}

@inproceedings{Donoghue:1995cz,
	author = "Donoghue, John F.",
	title = "{Introduction to the effective field theory description of gravity}",
	booktitle = "{Advanced School on Effective Theories}",
	eprint = "gr-qc/9512024",
	archivePrefix = "arXiv",
	reportNumber = "UMHEP-424",
	month = "6",
	year = "1995"
}

@article{Plenio2005,
	author = {Plenio, M. B.},
	doi = {10.1103/physrevlett.95.090503},
	issn = {1079-7114},
	journal = {Phys. Rev. Lett.},
	month = aug,
	number = {9},
	pages = {090503},
	publisher = {American Physical Society (APS)},
	title = {Logarithmic Negativity: A Full Entanglement Monotone That is not Convex},
	url = {http://dx.doi.org/10.1103/PhysRevLett.95.090503},
	volume = {95},
	year = {2005}}

@article{Carney:2018ofe,
	archiveprefix = {arXiv},
	author = {Carney, Daniel and Stamp, Philip C. E. and Taylor, Jacob M.},
	doi = {10.1088/1361-6382/aaf9ca},
	eprint = {1807.11494},
	journal = {Class. Quant. Grav.},
	number = {3},
	pages = {034001},
	primaryclass = {quant-ph},
	title = {{Tabletop experiments for quantum gravity: a user\textquoteright{}s manual}},
	volume = {36},
	year = {2019}}

@article{PhysRevA.101.063804,
	author = {Miao, Haixing and Martynov, Denis and Yang, Huan and Datta, Animesh},
	doi = {10.1103/PhysRevA.101.063804},
	issue = {6},
	journal = {Phys. Rev. A},
	month = {Jun},
	numpages = {7},
	pages = {063804},
	publisher = {American Physical Society},
	title = {Quantum correlations of light mediated by gravity},
	url = {https://link.aps.org/doi/10.1103/PhysRevA.101.063804},
	volume = {101},
	year = {2020}}

@article{liu2021gravitational,
	author = {Liu, Yulong and Mummery, Jay and Zhou, Jingwei and Sillanp{\"a}{\"a}, Mika A},
	journal = {Phys. Rev. Applied},
	number = {3},
	pages = {034004},
	publisher = {APS},
	title = {Gravitational Forces Between Nonclassical Mechanical Oscillators},
	volume = {15},
	year = {2021}}

@article{kafri2015bounds,
	author = {Kafri, D and Milburn, GJ and Taylor, JM},
	journal = {New Journal of Physics},
	number = {1},
	pages = {015006},
	publisher = {IOP Publishing},
	title = {Bounds on quantum communication via Newtonian gravity},
	volume = {17},
	year = {2015}}

@article{kafri2013noise,
	archiveprefix = {arXiv},
	author = {Kafri, Dvir and Taylor, JM},
	journal = {arXiv e-prints},
	eprint = {1311.4558},
	primaryclass = {quant-ph},
	title = {A noise inequality for classical forces},
	year = {2013}}

@article{Bahrami:2015wma,
	archiveprefix = {arXiv},
	author = {Bahrami, M. and Bassi, A. and McMillen, S. and Paternostro, M. and Ulbricht, H.},
	journal = {arXiv e-prints},
	eprint = {1507.05733},
	month = {7},
	primaryclass = {quant-ph},
	title = {{Is Gravity Quantum?}},
	year = {2015}}

@article{Carlesso:2019cuh,
	archiveprefix = {arXiv},
	author = {Carlesso, Matteo and Bassi, Angelo and Paternostro, Mauro and Ulbricht, Hendrik},
	doi = {10.1088/1367-2630/ab41c1},
	eprint = {1906.04513},
	journal = {New J. Phys.},
	number = {9},
	pages = {093052},
	primaryclass = {quant-ph},
	title = {{Testing the gravitational field generated by a quantum superposition}},
	volume = {21},
	year = {2019}}

@article{Matsumura:2020law,
	archiveprefix = {arXiv},
	author = {Matsumura, Akira and Yamamoto, Kazuhiro},
	doi = {10.1103/PhysRevD.102.106021},
	eprint = {2010.05161},
	journal = {Phys. Rev. D},
	number = {10},
	pages = {106021},
	primaryclass = {gr-qc},
	title = {{Gravity-induced entanglement in optomechanical systems}},
	volume = {102},
	year = {2020}}

@article{anastopoulos2015probing,
	author = {Anastopoulos, Charis and Hu, Bei-Lok},
	journal = {Class. Quantum Gravity},
	number = {16},
	pages = {165022},
	publisher = {IOP Publishing},
	title = {Probing a gravitational cat state},
	volume = {32},
	year = {2015}}

@article{Qvarfort:2018uag,
	archiveprefix = {arXiv},
	author = {Qvarfort, Sofia and Bose, Sougato and Serafini, Alessio},
	doi = {10.1088/1361-6455/abbe8d},
	eprint = {1812.09776},
	journal = {J. Phys. B},
	number = {23},
	pages = {235501},
	primaryclass = {quant-ph},
	title = {{Mesoscopic entanglement through central\textendash{}potential interactions}},
	volume = {53},
	year = {2020}}

@article{haine2021searching,
	author = {Haine, Simon A},
	journal = {New J. Phys.},
	number = {3},
	pages = {033020},
	publisher = {IOP Publishing},
	title = {Searching for signatures of quantum gravity in quantum gases},
	volume = {23},
	year = {2021}}

@article{howl2021non,
	author = {Howl, Richard and Vedral, Vlatko and Naik, Devang and Christodoulou, Marios and Rovelli, Carlo and Iyer, Aditya},
	journal = {PRX Quantum},
	number = {1},
	pages = {010325},
	publisher = {APS},
	title = {Non-Gaussianity as a signature of a quantum theory of gravity},
	volume = {2},
	year = {2021}}

@article{bose2017spin,
	author = {Bose, Sougato and Mazumdar, Anupam and Morley, Gavin W and Ulbricht, Hendrik and Toro{\v{s}}, Marko and Paternostro, Mauro and Geraci, Andrew A and Barker, Peter F and Kim, MS and Milburn, Gerard},
	journal = {Phys. Rev. Lett.},
	number = {24},
	pages = {240401},
	publisher = {APS},
	title = {Spin Entanglement Witness for Quantum Gravity},
	volume = {119},
	year = {2017}}

@article{marletto2017gravitationally,
	author = {Marletto, Chiara and Vedral, Vlatko},
	journal = {Phys. Rev. Lett.},
	number = {24},
	pages = {240402},
	publisher = {APS},
	title = {Gravitationally induced entanglement between two massive particles is sufficient evidence of quantum effects in gravity},
	volume = {119},
	year = {2017}}

@article{Pedernales:2021dja,
	archiveprefix = {arXiv},
	author = {Pedernales, Julen S. and Streltsov, Kirill and Plenio, Martin B.},
	doi = {10.1103/PhysRevLett.128.110401},
	eprint = {2104.14524},
	journal = {Phys. Rev. Lett.},
	number = {11},
	pages = {110401},
	primaryclass = {quant-ph},
	title = {{Enhancing Gravitational Interaction between Quantum Systems by a Massive Mediator}},
	volume = {128},
	year = {2022}}

@article{Datta:2021ywm,
	archiveprefix = {arXiv},
	author = {Datta, Animesh and Miao, Haixing},
	journal = {Quantum Science and Technology},
	eprint = {2104.04414},
	month = {4},
	primaryclass = {gr-qc},
	title = {{Signatures of the quantum nature of gravity in the differential motion of two masses}},
	year = {2021}}

@article{Carney:2021vvt,
	archiveprefix = {arXiv},
	author = {Carney, Daniel},
	doi = {10.1103/PhysRevD.105.024029},
	eprint = {2108.06320},
	journal = {Phys. Rev. D},
	number = {2},
	pages = {024029},
	primaryclass = {quant-ph},
	title = {{Newton, entanglement, and the graviton}},
	volume = {105},
	year = {2022}}

@article{Kafri:2014zsa,
	archiveprefix = {arXiv},
	author = {Kafri, D. and Taylor, J. M. and Milburn, G. J.},
	doi = {10.1088/1367-2630/16/6/065020},
	eprint = {1401.0946},
	journal = {New J. Phys.},
	pages = {065020},
	primaryclass = {quant-ph},
	title = {{A classical channel model for gravitational decoherence}},
	volume = {16},
	year = {2014}}

@article{Hall:2017nzl,
	archiveprefix = {arXiv},
	author = {Hall, Michael J. W. and Reginatto, Marcel},
	doi = {10.1088/1751-8121/aaa734},
	eprint = {1707.07974},
	journal = {J. Phys. A},
	number = {8},
	pages = {085303},
	primaryclass = {quant-ph},
	title = {{On two recent proposals for witnessing nonclassical gravity}},
	volume = {51},
	year = {2018}}

@article{martin2023gravity,
	author = {Mart{\'\i}n-Mart{\'\i}nez, Eduardo and Perche, T Rick},
	journal = {Physical Review D},
	number = {10},
	pages = {L101702},
	publisher = {APS},
	title = {What gravity mediated entanglement can really tell us about quantum gravity},
	volume = {108},
	year = {2023}}

@article{Trillo:2024kqz,
	archiveprefix = {arXiv},
	author = {Trillo, David and Navascu\'es, Miguel},
	journal = {Phys. Rev. D},
	eprint = {2411.02287},
	month = {11},
	primaryclass = {quant-ph},
	title = {{The Di\'osi-Penrose model of classical gravity predicts gravitationally induced entanglement}},
	year = {2024}}

@article{Horodecki:2009zz,
	archiveprefix = {arXiv},
	author = {Horodecki, Ryszard and Horodecki, Pawel and Horodecki, Michal and Horodecki, Karol},
	doi = {10.1103/RevModPhys.81.865},
	eprint = {quant-ph/0702225},
	journal = {Rev. Mod. Phys.},
	pages = {865--942},
	title = {{Quantum entanglement}},
	volume = {81},
	year = {2009}}

@article{Mikheev:1987qk,
	author = {Mikheev, S. P. and Smirnov, A. Yu.},
	doi = {10.1070/PU1987v030n09ABEH002961},
	journal = {Sov. Phys. Usp.},
	pages = {759--790},
	title = {{Resonance Oscillations of Neutrinos in Matter}},
	volume = {30},
	year = {1987}}

@article{Kosowsky:1994cy,
	archiveprefix = {arXiv},
	author = {Kosowsky, Arthur},
	doi = {10.1006/aphy.1996.0020},
	eprint = {astro-ph/9501045},
	journal = {Annals Phys.},
	pages = {49--85},
	reportnumber = {FERMILAB-PUB-94-206-A},
	title = {{Cosmic microwave background polarization}},
	volume = {246},
	year = {1996}}

@article{Bavarsad:2009hm,
	archiveprefix = {arXiv},
	author = {Bavarsad, E. and Haghighat, M. and Rezaei, Z. and Mohammadi, R. and Motie, I. and Zarei, M.},
	doi = {10.1103/PhysRevD.81.084035},
	eprint = {0912.2993},
	journal = {Phys. Rev. D},
	pages = {084035},
	primaryclass = {hep-th},
	title = {{Generation of circular polarization of the CMB}},
	volume = {81},
	year = {2010}}

@article{Bartolo:2018igk,
	archiveprefix = {arXiv},
	author = {Bartolo, Nicola and Hoseinpour, Ahmad and Orlando, Giorgio and Matarrese, Sabino and Zarei, Moslem},
	doi = {10.1103/PhysRevD.98.023518},
	eprint = {1804.06298},
	journal = {Phys. Rev. D},
	number = {2},
	pages = {023518},
	primaryclass = {gr-qc},
	title = {{Photon-graviton scattering: A new way to detect anisotropic gravitational waves?}},
	volume = {98},
	year = {2018}}

@article{Bartolo:2019eac,
	archiveprefix = {arXiv},
	author = {Bartolo, Nicola and Hoseinpour, Ahmad and Matarrese, Sabino and Orlando, Giorgio and Zarei, Moslem},
	doi = {10.1103/PhysRevD.100.043516},
	eprint = {1903.04578},
	journal = {Phys. Rev. D},
	number = {4},
	pages = {043516},
	primaryclass = {hep-ph},
	title = {{CMB Circular and B-mode Polarization from New Interactions}},
	volume = {100},
	year = {2019}}

@article{Hoseinpour:2020hic,
	archiveprefix = {arXiv},
	author = {Hoseinpour, Ahmad and Zarei, Moslem and Orlando, Giorgio and Bartolo, Nicola and Matarrese, Sabino},
	doi = {10.1103/PhysRevD.102.063501},
	eprint = {2006.14418},
	journal = {Phys. Rev. D},
	number = {6},
	pages = {063501},
	primaryclass = {hep-ph},
	title = {{CMB $V$ modes from photon-photon forward scattering revisited}},
	volume = {102},
	year = {2020}}

@article{Chen:2024xvm,
 	author = "Chen, Lin-Qing and Giacomini, Flaminia",
 	title = "{Quantum effects in gravity beyond the Newton potential from a delocalised quantum source}",
 	eprint = "2402.10288",
 	archivePrefix = "arXiv",
 	primaryClass = "quant-ph",
 	month = "2",
 	year = "2024",
 	journal = {arXiv e-prints}
 }

@article{Belenchia2018,
	adsurl = {https://ui.adsabs.harvard.edu/abs/2018PhRvD..98l6009B},
	archiveprefix = {arXiv},
	author = {{Belenchia}, Alessio and {Wald}, Robert M. and {Giacomini}, Flaminia and {Castro-Ruiz}, Esteban and {Brukner}, {\v{C}}aslav and {Aspelmeyer}, Markus},
	doi = {10.1103/PhysRevD.98.126009},
	eid = {126009},
	eprint = {1807.07015},
	journal = {\prd},
	month = dec,
	number = {12},
	pages = {126009},
	primaryclass = {quant-ph},
	title = {{Quantum superposition of massive objects and the quantization of gravity}},
	volume = {98},
	year = 2018}

@article{Bose:2017nin,
	adsurl = {https://ui.adsabs.harvard.edu/abs/2017PhRvL.119x0401B},
	archiveprefix = {arXiv},
	author = {{Bose}, Sougato and {Mazumdar}, Anupam and {Morley}, Gavin W. and {Ulbricht}, Hendrik and {Toro{\v{s}}}, Marko and {Paternostro}, Mauro and {Geraci}, Andrew A. and {Barker}, Peter F. and {Kim}, M.~S. and {Milburn}, Gerard},
	doi = {10.1103/PhysRevLett.119.240401},
	eid = {240401},
	eprint = {1707.06050},
	journal = {\prl},
	month = dec,
	number = {24},
	pages = {240401},
	primaryclass = {quant-ph},
	title = {{Spin Entanglement Witness for Quantum Gravity}},
	volume = {119},
	year = 2017}

@article{Marletto:2017kzi,
	adsurl = {https://ui.adsabs.harvard.edu/abs/2017PhRvL.119x0402M},
	archiveprefix = {arXiv},
	author = {{Marletto}, C. and {Vedral}, V.},
	doi = {10.1103/PhysRevLett.119.240402},
	eid = {240402},
	eprint = {1707.06036},
	journal = {\prl},
	month = dec,
	number = {24},
	pages = {240402},
	primaryclass = {quant-ph},
	title = {{Gravitationally Induced Entanglement between Two Massive Particles is Sufficient Evidence of Quantum Effects in Gravity}},
	volume = {119},
	year = 2017}

@article{Krisnanda:2019glc,
	adsurl = {https://ui.adsabs.harvard.edu/abs/2020npjQI...6...12K},
	archiveprefix = {arXiv},
	author = {{Krisnanda}, Tanjung and {Tham}, Guo Yao and {Paternostro}, Mauro and {Paterek}, Tomasz},
	doi = {10.1038/s41534-020-0243-y},
	eid = {12},
	eprint = {1906.08808},
	journal = {npj Quantum Information},
	month = jan,
	pages = {12},
	primaryclass = {quant-ph},
	title = {{Observable quantum entanglement due to gravity}},
	volume = {6},
	year = 2020}

@article{vandeKamp:2020rqh,
	archiveprefix = {arXiv},
	author = {van de Kamp, Thomas W. and Marshman, Ryan J. and Bose, Sougato and Mazumdar, Anupam},
	doi = {10.1103/PhysRevA.102.062807},
	eprint = {2006.06931},
	journal = {Phys. Rev. A},
	number = {6},
	pages = {062807},
	primaryclass = {quant-ph},
	title = {{Quantum Gravity Witness via Entanglement of Masses: Casimir Screening}},
	volume = {102},
	year = {2020}}

@article{Tilly.Qudits.2021,
	author = {Tilly, Jules and Marshman, Ryan J. and Mazumdar, Anupam and Bose, Sougato},
	doi = {10.1103/PhysRevA.104.052416},
	issue = {5},
	journal = {Phys. Rev. A},
	month = {Nov},
	numpages = {13},
	pages = {052416},
	publisher = {American Physical Society},
	title = {{Qudits for witnessing quantum-gravity-induced entanglement of masses under decoherence}},
	url = {https://link.aps.org/doi/10.1103/PhysRevA.104.052416},
	volume = {104},
	year = {2021}}

@article{Schut:2021svd,
	archiveprefix = {arXiv},
	author = {Schut, Martine and Tilly, Jules and Marshman, Ryan J. and Bose, Sougato and Mazumdar, Anupam},
	doi = {10.1103/PhysRevA.105.032411},
	eprint = {2110.14695},
	journal = {Phys. Rev. A},
	number = {3},
	pages = {032411},
	primaryclass = {quant-ph},
	title = {{Improving resilience of quantum-gravity-induced entanglement of masses to decoherence using three superpositions}},
	volume = {105},
	year = {2022}}

@article{Kent:2021ljj,
	archiveprefix = {arXiv},
	author = {Kent, Adrian and Pital\'ua-Garc\'\i{}a, Dami\'an},
	doi = {10.1103/PhysRevD.104.126030},
	eprint = {2109.02616},
	journal = {Phys. Rev. D},
	number = {12},
	pages = {126030},
	primaryclass = {quant-ph},
	title = {{Testing the nonclassicality of spacetime: What can we learn from Bell\textendash{}Bose et~al.-Marletto-Vedral experiments?}},
	volume = {104},
	year = {2021}}

@article{Bassi:2017szd,
	archiveprefix = {arXiv},
	author = {Bassi, Angelo and Gro\ss{}ardt, Andr\'e and Ulbricht, Hendrik},
	doi = {10.1088/1361-6382/aa864f},
	eprint = {1706.05677},
	journal = {Class. Quant. Grav.},
	number = {19},
	pages = {193002},
	primaryclass = {quant-ph},
	title = {{Gravitational Decoherence}},
	volume = {34},
	year = {2017}}

@article{Chevalier:2020uvv,
	archiveprefix = {arXiv},
	author = {Chevalier, Hadrien and Paige, A. J. and Kim, M. S.},
	doi = {10.1103/PhysRevA.102.022428},
	eprint = {2005.13922},
	journal = {Phys. Rev. A},
	number = {2},
	pages = {022428},
	primaryclass = {quant-ph},
	title = {{Witnessing the nonclassical nature of gravity in the presence of unknown interactions}},
	volume = {102},
	year = {2020}}

@article{Marletto:2024ltk,
	archiveprefix = {arXiv},
	author = {Marletto, Chiara and Vedral, Vlatko},
	journal = {arXiv e-prints},
	eprint = {2410.07262},
	month = {10},
	primaryclass = {quant-ph},
	title = {{Quantum-information methods for quantum gravity laboratory-based tests}},
	year = {2024}}

@article{Ma2021,
	author = {X. Ma and J. J. Viennot and S. Kotler and J. D. Teufel and K. W. Lehnert},
	doi = {10.1038/s41567-020-01102-1},
	journal = {Nature Physics},
	pages = {322},
	title = {Non-classical energy squeezing of a macroscopic mechanical oscillator},
	volume = {17},
	year = {2021}}

@article{Bild2023,
	author = {Marius Bild and Matteo Fadel and Yu Yang and Uwe von L{\"u}pke and Phillip Martin and Alessandro Bruno and Yiwen Chu},
	doi = {10.1126/science.adf7553},
	issue = {6642},
	journal = {Science},
	pages = {274},
	title = {Schr{\"o}dinger cat states of a 16-microgram mechanical oscillator},
	volume = {380},
	year = {2023}}

@article{Anastopoulos:2020cdp,
	archiveprefix = {arXiv},
	author = {Anastopoulos, Charis and Hu, Bei-Lok},
	doi = {10.1088/1361-6382/abbe6f},
	eprint = {2007.06446},
	journal = {Class. Quant. Grav.},
	number = {23},
	pages = {235012},
	primaryclass = {quant-ph},
	title = {{Quantum Superposition of Two Gravitational Cat States}},
	volume = {37},
	year = {2020}}

@article{Anastopoulos:2018drh,
	archiveprefix = {arXiv},
	author = {Anastopoulos, C. and Hu, Bei-Lok},
	journal = {arXiv e-prints},
	eprint = {1804.11315},
	month = {4},
	primaryclass = {quant-ph},
	title = {{Comment on ''A Spin Entanglement Witness for Quantum Gravity'' and on ''Gravitationally Induced Entanglement between Two Massive Particles is Sufficient Evidence of Quantum Effects in Gravity''}},
	year = {2018}}

@article{Vicentini:2024noi,
	archiveprefix = {arXiv},
	author = {Vicentini, Marta and Bernardi, Ettore and Moreva, Ekaterina and Piacentini, Fabrizio and Napoli, Carmine and Degiovanni, Ivo Pietro and Manzin, Alessandra and Genovese, Marco},
	journal = {arXiv e-prints},
	eprint = {2405.21029},
	month = {5},
	primaryclass = {quant-ph},
	title = {{Table-top nanodiamond interferometer enabling quantum gravity tests}},
	year = {2024}}

@article{Sugiyama:2022ixw,
	archiveprefix = {arXiv},
	author = {Sugiyama, Yuuki and Matsumura, Akira and Yamamoto, Kazuhiro},
	doi = {10.1103/PhysRevD.106.045009},
	eprint = {2203.09011},
	journal = {Phys. Rev. D},
	number = {4},
	pages = {045009},
	primaryclass = {quant-ph},
	title = {{Effects of photon field on entanglement generation in charged particles}},
	volume = {106},
	year = {2022}}

@article{Hidaka:2022gsv,
	archiveprefix = {arXiv},
	author = {Hidaka, Yoshimasa and Iso, Satoshi and Shimada, Kengo},
	doi = {10.1103/PhysRevD.107.085003},
	eprint = {2211.09441},
	journal = {Phys. Rev. D},
	number = {8},
	pages = {085003},
	primaryclass = {quant-ph},
	reportnumber = {KEK-TH-2453, YITP-22-136, RIKEN-iTHEMS-Report-22},
	title = {{Entanglement generation and decoherence in a two-qubit system mediated by relativistic quantum field}},
	volume = {107},
	year = {2023}}

@article{Gunnink.Gravitational.2023,
	author = {Fabian Gunnink and Anupam Mazumdar and Martine Schut and Marko Toro{\v s}},
	doi = {10.1088/1361-6382/ad0496},
	journal = {Classical and Quantum Gravity},
	month = {nov},
	number = {23},
	pages = {235006},
	publisher = {IOP Publishing},
	title = {Gravitational decoherence by the apparatus in the quantum-gravity-induced entanglement of masses},
	url = {https://dx.doi.org/10.1088/1361-6382/ad0496},
	volume = {40},
	year = {2023}}

@article{Rijavec:2020qxd,
	archiveprefix = {arXiv},
	author = {Rijavec, Simone and Carlesso, Matteo and Bassi, Angelo and Vedral, Vlatko and Marletto, Chiara},
	doi = {10.1088/1367-2630/abf3eb},
	eprint = {2012.06230},
	journal = {New J. Phys.},
	number = {4},
	pages = {043040},
	primaryclass = {quant-ph},
	title = {{Decoherence effects in non-classicality tests of gravity}},
	volume = {23},
	year = {2021}}

@article{Bose:2023gwh,
	archiveprefix = {arXiv},
	author = {Bose, Sougato and Fuentes, Ivette and Geraci, Andrew A. and Khan, Saba Mehsar and Qvarfort, Sofia and Rademacher, Markus and Rashid, Muddassar and Toro\v{s}, Marko and Ulbricht, Hendrik and Wanjura, Clara C.},
	journal = {Rev. Mod. Phys.},
	eprint = {2311.09218},
	month = {11},
	primaryclass = {quant-ph},
	title = {{Massive quantum systems as interfaces of quantum mechanics and gravity}},
	year = {2023}}

@book{Breuer2002,
	author = {Breuer, Heinz-Peter and Petruccione, Francesco},
	publisher = {Oxford University Press, USA},
	title = {The Theory of Open Quantum Systems},
	year = {2002}}

@book{Nielsen,
	author = {Nielsen, Michael A. and Chuang, Isaac L.},
	title = {Quantum Computation
	and Quantum Information},
	publisher = {Cambridge University Press},
	year = {2000}}

@book{Gerry,
	author = {C. C. Gerry and P. L. Knight },
	title = {Introductory Quantum Optics},
	publisher = {Cambridge University Press},
	year = {2005}}

@article{Christodoulou2019,
	author = {Christodoulou, Marios and Rovelli, Carlo},
	journal = {Physics Letters B},
	pages = {64--68},
	title = {On the possibility of laboratory evidence for quantum superposition of geometries},
	volume = {792},
	year = {2019}}

@article{danielson2022gravitationally,
	author = {Danielson, Daine L and Satishchandran, Gautam and Wald, Robert M},
	journal = {Physical Review D},
	number = {8},
	pages = {086001},
	publisher = {APS},
	title = {Gravitationally mediated entanglement: Newtonian field versus gravitons},
	volume = {105},
	year = {2022}}

@article{galley2022no,
	author = {Galley, Thomas D and Giacomini, Flaminia and Selby, John H},
	journal = {Quantum},
	pages = {779},
	publisher = {Verein zur F{\"o}rderung des Open Access Publizierens in den Quantenwissenschaften},
	title = {A no-go theorem on the nature of the gravitational field beyond quantum theory},
	volume = {6},
	year = {2022}}

@article{christodoulou2023locally,
	author = {Christodoulou, Marios and Di Biagio, Andrea and Aspelmeyer, Markus and Brukner, {\v{C}}aslav and Rovelli, Carlo and Howl, Richard},
	journal = {Physical Review Letters},
	number = {10},
	pages = {100202},
	publisher = {APS},
	title = {Locally mediated entanglement in linearized quantum gravity},
	volume = {130},
	year = {2023}}

@article{Toros:2020dbf,
	archiveprefix = {arXiv},
	author = {Toro\v{s}, Marko and Van De Kamp, Thomas W. and Marshman, Ryan J. and Kim, M. S. and Mazumdar, Anupam and Bose, Sougato},
	doi = {10.1103/PhysRevResearch.3.023178},
	eprint = {2007.15029},
	journal = {Phys. Rev. Res.},
	number = {2},
	pages = {023178},
	primaryclass = {gr-qc},
	title = {{Relative acceleration noise mitigation for nanocrystal matter-wave interferometry: Applications to entangling masses via quantum gravity}},
	volume = {3},
	year = {2021}}

@article{Wood:2021icq,
	archiveprefix = {arXiv},
	author = {Wood, B. D. and Bose, S. and Morley, G. W.},
	doi = {10.1103/PhysRevA.105.012824},
	eprint = {2105.02105},
	journal = {Phys. Rev. A},
	number = {1},
	pages = {012824},
	primaryclass = {quant-ph},
	title = {{Spin dynamical decoupling for generating macroscopic superpositions of a free-falling nanodiamond}},
	volume = {105},
	year = {2022}}

@article{Wan_2016,
	author = {Wan, C. and Scala, M. and Morley, G. W. and Rahman, ATM. A. and Ulbricht, H. and Bateman, J. and Barker, P. F. and Bose, S. and Kim, M. S.},
	doi = {10.1103/physrevlett.117.143003},
	issn = {1079-7114},
	journal = {Physical Review Letters},
	month = sep,
	number = {14},
	publisher = {American Physical Society (APS)},
	title = {Free Nano-Object Ramsey Interferometry for Large Quantum Superpositions},
	volume = {117},
	year = {2016}}

@article{Abrahao:2023lle,
	archiveprefix = {arXiv},
	author = {Abrahao, Luca and Coradeschi, Francesco and Frassino, Antonia Micol and Guerreiro, Thiago and Rittenhouse West, Jennifer and Schioppa, Junior., Enrico},
	doi = {10.1088/1361-6382/ad134e},
	eprint = {2308.12713},
	journal = {Class. Quant. Grav.},
	number = {1},
	pages = {015029},
	primaryclass = {gr-qc},
	title = {{The quantum optics of gravitational waves}},
	volume = {41},
	year = {2024}}

@article{Arani:2023ljw,
	archiveprefix = {arXiv},
	author = {Arani, Fateme Shojaei and Harouni, Malek Bagheri and Lamine, Brahim and Blanchard, Alain},
	journal = {arXiv e-prints},
	eprint = {2312.00474},
	month = {12},
	primaryclass = {gr-qc},
	title = {{Constraining tensor-to-scalar ratio based on VLBI observations: PGWs induced-incoherence approach}},
	year = {2023}}

@article{Colas:2024xjy,
	archiveprefix = {arXiv},
	author = {Colas, Thomas and de Rham, Claudia and Kaplanek, Greg},
	doi = {10.1088/1475-7516/2024/05/025},
	eprint = {2401.02832},
	journal = {JCAP},
	pages = {025},
	primaryclass = {hep-th},
	reportnumber = {Imperial/TP/2024/CdR/01; Imperial/TP/2024/GK/01},
	title = {{Decoherence out of fire: purity loss in expanding and contracting universes}},
	volume = {05},
	year = {2024}}

@article{Sugiyama:2024equ,
	archiveprefix = {arXiv},
	author = {Sugiyama, Yuuki and Matsumura, Akira and Yamamoto, Kazuhiro},
	doi = {10.1103/PhysRevD.110.045016},
	eprint = {2401.03867},
	journal = {Phys. Rev. D},
	number = {4},
	pages = {045016},
	primaryclass = {quant-ph},
	title = {{Quantumness of the gravitational field: A perspective on monogamy relation}},
	volume = {110},
	year = {2024}}

@article{Gallock-Yoshimura:2024wpz,
	archiveprefix = {arXiv},
	author = {Gallock-Yoshimura, Kensuke and Sugiyama, Yuuki and Matsumura, Akira and Yamamoto, Kazuhiro},
	doi = {10.1103/PhysRevD.110.085018},
	eprint = {2407.14581},
	journal = {Phys. Rev. D},
	number = {8},
	pages = {085018},
	primaryclass = {quant-ph},
	title = {{Decoherence of spin superposition state caused by a quantum electromagnetic field}},
	volume = {110},
	year = {2024}}

@article{Anastopoulos:2022vvb,
	archiveprefix = {arXiv},
	author = {Anastopoulos, Charis and Hu, Bei-Lok},
	doi = {10.3390/e24040490},
	eprint = {2202.02789},
	journal = {Entropy},
	number = {4},
	pages = {490},
	primaryclass = {quant-ph},
	title = {{Gravity, Quantum Fields and Quantum Information: Problems with Classical Channel and Stochastic Theories}},
	volume = {24},
	year = {2022}}

@article{Sharifian:2025olk,
	archiveprefix = {arXiv},
	author = {Sharifian, Mohammad and Zarei, Moslem and Bartolo, Nicola and Matarrese, Sabino},
	journal = {arXiv e-prints},
	eprint = {2501.13678},
	month = {1},
	primaryclass = {gr-qc},
	title = {{Exploring gravitational impulse via quantum Boltzmann equation}},
	year = {2025}}

@article{Martin-Martinez:2022uio,
	author = "Mart\'\i{}n-Mart\'\i{}nez, Eduardo and Perche, T. Rick",
	title = "{What gravity mediated entanglement can really tell us about quantum gravity}",
	eprint = "2208.09489",
	archivePrefix = "arXiv",
	primaryClass = "quant-ph",
	doi = "10.1103/PhysRevD.108.L101702",
	journal = "Phys. Rev. D",
	volume = "108",
	number = "10",
	pages = "L101702",
	year = "2023"
}

@article{Reginatto:2018ksa,
 	author = "Reginatto, Marcel and Hall, Michael J. W.",
 	title = "{Entanglement of quantum fields via classical gravity}",
 	eprint = "1809.04989",
 	archivePrefix = "arXiv",
 	primaryClass = "gr-qc",
 	month = "9",
 	year = "2018",
 	journal = {arXiv e-prints}
 }

@article{Sigl:1993ctk,
	author = {Sigl, G. and Raffelt, G.},
	doi = {10.1016/0550-3213(93)90175-O},
	journal = {Nucl. Phys. B},
	pages = {423--451},
	reportnumber = {MPI-PH-92-112},
	title = {{General kinetic description of relativistic mixed neutrinos}},
	volume = {406},
	year = {1993}}

@article{Zarei:2021dpb,
	author = "Zarei, Moslem and Bartolo, Nicola and Bertacca, Daniele and Ricciardone, Angelo and Matarrese, Sabino",
	title = "{Non-Markovian open quantum system approach to the early Universe: Damping of gravitational waves by matter}",
	eprint = "2104.04836",
	archivePrefix = "arXiv",
	primaryClass = "astro-ph.CO",
	doi = "10.1103/PhysRevD.104.083508",
	journal = "Phys. Rev. D",
	volume = "104",
	number = "8",
	pages = "083508",
	year = "2021"
}

@article{Sharifian:2023jem,
	archiveprefix = {arXiv},
	author = {Sharifian, Mohammad and Zarei, Moslem and Abdi, Mehdi and Bartolo, Nicola and Matarrese, Sabino},
	doi = {10.1103/PhysRevD.109.043510},
	eprint = {2309.07236},
	journal = {Phys. Rev. D},
	number = {4},
	pages = {043510},
	primaryclass = {gr-qc},
	title = {{Open quantum system approach to the gravitational decoherence of spin-1/2 particles}},
	volume = {109},
	year = {2024}}

@article{Aziz:2025ypo,
	author = "Aziz, Joseph and Howl, Richard",
	title = "{Classical theories of gravity produce entanglement}",
	eprint = "2510.19714",
	archivePrefix = "arXiv",
	primaryClass = "quant-ph",
	doi = "10.1038/s41586-025-09595-7",
	journal = "Nature",
	volume = "646",
	number = "8086",
	pages = "813--817",
	year = "2025"
}

\clearpage
\onecolumngrid
%\begin{center}
%	\textbf{\large Supplementary Material}
%\end{center}

% \setcounter{equation}{0}
% \setcounter{figure}{0}
% \setcounter{table}{0}
% \setcounter{page}{1}
% \makeatletter
% \renewcommand*{\thesection}{S\arabic{section}}
% \renewcommand*{\theequation}{S\arabic{equation}}
% \renewcommand*{\thefigure}{S\arabic{figure}}

\appendix

\section{  Review of spin entanglement  due to gravitational interaction}\label{appendix:A}
We first consider a simplified version of the theory presented in the main text to clarify how the states of two neutral test masses $1$ and $2$, with masses $m_1$ and $m_2$, respectively, become entangled through gravitational interaction. Each mass with embeded spin is held in a steady superposition of two spatially separated states $|L\rangle$ and $|R\rangle$, as shown in Fig. \eqref{1}, for a time interval $\tau$. Later, we will describe how this setup can be realized, following the scheme depicted in Fig. \eqref{1} \cite{Bose:2017nin,Marletto:2017kzi}.
Assume the centers of $|L\rangle$ and $|R\rangle$ are separated by a distance $\Delta x$. Each of these states is represented by localized Gaussian wavepackets with widths much smaller than $\Delta x$, ensuring that $\langle L|R \rangle = 0$. Furthermore, let there be a separation $d$ between the centers of the superpositions, such that even at the closest approach of the masses ($d - \Delta x$), short-range interactions such as the Casimir-Polder force are shielded.

Due to the spatial separation, distinct components of the superposition have different gravitational interaction energies, as the distances between the masses vary. This leads to different rates of phase evolution. Under these circumstances, the time evolution of the joint quantum state of the two masses is determined solely by their mutual gravitational interaction.
Initially, each mass is prepared in a superposition state, hence the total initial state is given by \cite{Bose:2017nin}
%\begin{eqnarray}
%|\psi(t = 0)\rangle_{12} = \frac{1}{2} \left( |L\rangle_1 + |R\rangle_1 \right)
%\otimes 
%\left( |L\rangle_2 + |R\rangle_2 \right)~.
%\end{eqnarray}
%The density matrix
%Expanding this, the state becomes
\begin{eqnarray}
	|\psi(t = 0)\rangle_{12} = \frac{1}{2} \Big( |L\rangle_1 |L\rangle_2 + |L\rangle_1 |R\rangle_2 
	+ |R\rangle_1 |L\rangle_2 + |R\rangle_1 |R\rangle_2 \Big)~.
\end{eqnarray}
It is straightforward to show that after a time $\tau$, the state evolves into
\begin{eqnarray}
	|\psi(t = \tau)\rangle_{12} &=& \frac{e^{i\phi}}{\sqrt{2}} \Bigg\{ |L\rangle_1 \frac{1}{\sqrt{2}} 
	\left( |L\rangle_2 + e^{i\Delta \phi_{LR}} |R\rangle_2 \right) \nonumber \\
	&& + |R\rangle_1 \frac{1}{\sqrt{2}} 
	\left( e^{i\Delta \phi_{RL}} |L\rangle_2 + |R\rangle_2 \right) \Bigg\}~.
\end{eqnarray}
Here, the accumulated phases depend on the gravitational interaction of the masses and are given by
\begin{eqnarray}
	\Delta \phi_{RL} &=& \phi_{RL} - \phi, \quad \Delta \phi_{LR} = \phi_{LR} - \phi, \\
	\phi_{RL} &\approx& \frac{G m_1 m_2 \tau}{\hbar (d - \Delta x)}, \quad 
	\phi_{LR} \approx \frac{G m_1 m_2 \tau}{\hbar (d + \Delta x)}, \quad 
	\phi \approx \frac{G m_1 m_2 \tau}{\hbar d}~. \label{phi}
\end{eqnarray}
Each mass can be treated as an effective orbital qubit.
This reduces the problem to a two-qubit system, with the spatial degree of freedom encoding the qubit states.
For having entanglement between the qubits the total state must be inseparable.
In other words, the following states must not be identical up to a global phase.
\begin{eqnarray}
	\frac{1}{\sqrt{2}} \left( |L\rangle_2 + e^{i\Delta \phi_{LR}} |R\rangle_2 \right) \quad \text{and} \quad 
	\frac{1}{\sqrt{2}} \left( e^{i\Delta \phi_{RL}} |L\rangle_2 + |R\rangle_2 \right)~,
\end{eqnarray}
This occurs when $\Delta \phi_{LR} + \Delta \phi_{RL} \neq 2n\pi$, where $n$ is an integer. In this case, the state $|\psi(t = \tau)\rangle_{12}$ cannot be factorized, and the masses are entangled in their orbital states.

\section{ Significance of Entanglement}\label{appendix:B}
Witnessing this entanglement proves that a quantum field mediated the interaction, as a classical field cannot entangle spatially separated objects. The degree of entanglement increases monotonically with sum of the phases $\Delta \phi_{LR} + \Delta \phi_{RL}$, when evolving from $0$ to $2\pi$.
A maximally entangled state is formed when $\Delta \phi_{LR} + \Delta \phi_{RL}=\pi$.
Equivalently, to ensure significant entanglement, one must have
\begin{eqnarray}
	\Delta \phi_{LR} + \Delta \phi_{RL} \sim \mathcal{O}(1)~,
\end{eqnarray}
This highlights the importance of choosing parameters $d$, $\Delta x$, and $\tau$ such that the gravitational phase difference is appreciable and experimentally observable.

The density operator for a pure state $|\psi\rangle$ is defined as $\rho = |\psi\rangle \langle \psi|$. For the initial state
\begin{eqnarray}
	\rho(t = 0) = \frac{1}{4} 
	\begin{pmatrix}
		1 & 1 & 1 & 1 \\
		1 & 1 & 1 & 1 \\
		1 & 1 & 1 & 1 \\
		1 & 1 & 1 & 1
	\end{pmatrix}~, \label{initial}
\end{eqnarray}
in the basis $\{|LL\rangle, |LR\rangle, |RL\rangle, |RR\rangle\}$~.
After the evolution time $\tau$, the density matrix becoems
\begin{eqnarray}
	\rho(t = \tau) = \frac{1}{4} 
	\begin{pmatrix}
		1 & e^{-i\Delta \phi_{LR}} & e^{-i\Delta \phi_{RL}} & 1 \\
		e^{i\Delta \phi_{LR}} & 1 & e^{-i(\Delta \phi_{RL} - \Delta \phi_{LR})} & e^{i\Delta \phi_{LR}}  \\
		e^{i\Delta \phi_{RL}} & e^{i(\Delta \phi_{RL} - \Delta \phi_{LR})} & 1 & e^{i\Delta \phi_{LR}} \\
		1&e^{-i\Delta \phi_{RL}}& e^{-i\Delta \phi_{LR}} & 1
	\end{pmatrix}~.\label{density1}
\end{eqnarray}
By tracing out the subsystems $B$ yields the reduced density matrix for mass $A$ reads
\begin{eqnarray}
	\rho_A = \mathrm{Tr}_B(\rho) = 
	\frac{1}{2} 
	\begin{pmatrix}
		1 & \cos\frac{1}{2}\left(\Delta \phi_{LR} + \Delta \phi_{RL}\right) \\
		\cos\frac{1}{2}\left(\Delta \phi_{LR} + \Delta \phi_{RL}\right) & 1
	\end{pmatrix}~.
\end{eqnarray}
Any mixedness in this reduced density matrix signals an entanglement between the two particles, which occurs for $\Delta \phi_{LR} + \Delta \phi_{RL} \neq 2n\pi$, confirming entanglement between the masses.
The entanglement generated by gravitational interaction proves the quantum nature of the mediating field. Classical fields cannot induce entanglement.
%The degree of entanglement, quantified by measures such as concurrence or von Neumann entropy, depends on $\Delta \phi_{LR} + \Delta \phi_{RL}$. Maximal entanglement occurs when this phase sum approaches $\pi$, highlighting the importance of choosing experimental parameters to ensure observable quantum effects.

To verify that \(\rho(t = \tau)\) is entangled, we compute the negativity which is a standard monotone measure of entanglement.
To do so one first finds the partial transposition of the density matrix, \(\rho^{\Gamma}\).
The logarithmic negativity is then computed as \cite{Plenio2005}
\begin{equation}
	E_\mathcal{N}=\log_2\Vert\rho^\Gamma\Vert_1,
\end{equation}
where $\Vert \cdot \Vert_1$ is the trace norm.
\begin{figure}[tb]
	\includegraphics[width=0.4\columnwidth]{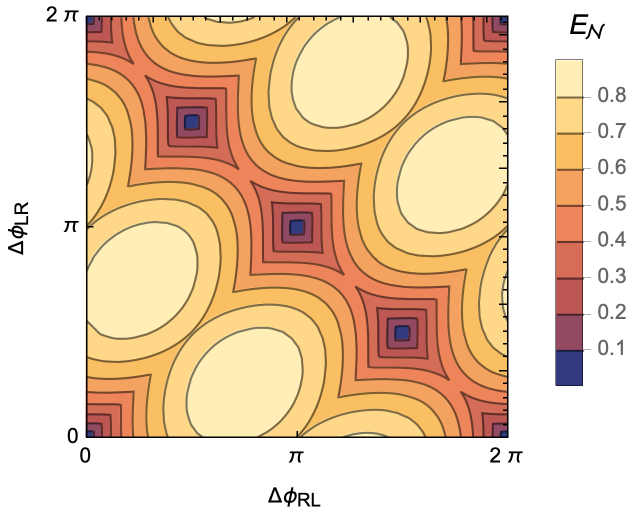}
	\caption{Logarithmic negativity as a function on the accumulated phases.}
	\label{fig:negativity}
\end{figure}
In fact, the entanglement between the massive orbital qubits increases monotonically over $\Delta \phi_{LR} + \Delta \phi_{RL}$ evolving from $0$ to $\pi$, with the entanglement being maximal for $\pi$. Our numerical investigations (Fig. \ref{fig:negativity}) show that this behavior holds for a range of parameters, confirming that the entanglement grows steadily as the sum of $\Delta \phi_{LR}$ and $\Delta \phi_{RL}$ approaches $\pi$. Thus, it is worth seeking conditions such that
\begin{eqnarray}
	\Delta \phi_{LR}  \approx 1-   \Delta \phi_{RL}~,
\end{eqnarray}
so that the entanglement between the states of the masses is significant and can be easily witnessed. Specifically, numerical simulations demonstrate that for values close to $\pi$, the entanglement becomes increasingly detectable, providing a clear signature of the interaction between the states.

\section{ Derivation of the density matrix for model I} \label{appendix:C}
By invoking the S-matrix formalism, we express the effective interaction Hamiltonian in Fourier space from Eq.~\eqref{Hmodel1}, taking the transverse-traceless limit, as follows
\begin{align}
	\hat{H}_{\inte}^{\rm (I)} &=- \frac{1}{4}\kappa^2 VV' \sum_{rr',ss'}
	\int d\mathbf{x} \, d\mathbf{x}' \, d\p \, d\p' \, d\q \, d\q' \, d\K \, \frac{p^0 q^0 }{|\K|^2}
	\delta^3_{\sigma_{0}}(\x-\bar{\x}_r) \delta^3_{\sigma'_{0}}(\x'-\bar{\x}_s') e^{i(\K+\p-\p')\cdot \x} e^{i(\q-\q'-\K)\cdot \x'}  \notag \\
	&\hspace{3cm} \times  	 
	\bar{u}_{s'}(\q') (I\otimes \sigma^3) u_s(\q) \bar{u}_{r'}(\p') (I\otimes \sigma^3) u_r(\p) b_{s'}^\dag(\q')b_{s}(\q)a_{r'}^\dag(\p')a_r(\p)~,
	\label{effH5}
\end{align}
where $I$ and $\sigma^3$ denote the identity and the third Pauli matrix, respectively. 
We now proceed by substituting the interaction Hamiltonian \eqref{effH5} into the forward scattering term—the first-order term in the interaction Hamiltonian—of the quantum Boltzmann equation \eqref{QBE1}. This substitution enables us to determine the evolution of $\rho_{IJ}$ up to the second order in $\kappa$. 
By integrating over $\mathbf{x}$ and $\mathbf{x}'$, we arrive at
\begin{align}
	\dot{\rho}_{IJ} &=- i \frac{\kappa^2}{4} \!\sum_{rr'}\!\sum_{ss'} \!\int \hspace{-1mm}d\p d\p' d\q d\q' d\K \,\frac{ip^0q^0}{|\K|^2} e^{-i(\K+\p-\p')\cdot \bar{\x}_r}
	e^{-i(\q-\q'-\K)\cdot \bar{\x}_s'} e^{-\sigma^2_0|\K+\p-\p'|^2}e^{-\sigma'^2_0|\K+\q'-\q|^2} 
	\notag \\
	&\hspace{3cm}
	\times  \bar{u}_{s'}(\q') (I\otimes \sigma^3) u_s(\q) \bar{u}_{r'}(\p') (I\otimes \sigma^3) u_r(\p) 
	\notag \\
	&\hspace{3cm}
	\times \Big[ \langle b_{s'}^\dag(\q') b_s(\q) a_{r'}^\dag(\p') a_r(\p) \hat{\mathcal{D}}_{IJ}(\K) \rangle 
	- \langle \hat{\mathcal{D}}_{IJ}(\K) b_{s'}^\dag(\q') b_s(\q) a_{r'}^\dag(\p') a_r(\p) \rangle \Big]~,
	\label{QBE3-1}
\end{align}
where we have used $V,V' = (2\pi)^3 \delta^3(0)$.
The wave packet form of particles with finite widths $\sigma_0$ and $\sigma'_0$ introduces an effective ultraviolet cutoff for $|\mathbf{K}|$. We exploit the independence of the spin Hilbert spaces to express the number operators as $\hat{\mathcal{D}}_{IJ} = a_i^\dag a_j \otimes b_k^\dag b_l$, where we have folded the density matrix indices such that ${i,j,k,l} = 1,2$. 
Substituting this into Eq.~\eqref{QBE3-1}, we arrive at
\begin{align}
	\dot{\rho}_{IJ}\! &=\!-\frac{i}{4}  \kappa^2\hspace{-1.5mm}\sum_{rr'ss'}\hspace{-1mm}  \int\! d\p d\p' \, d\q d\q' \, d\K \, \frac{p^0q^0}{|\K|^2} e^{-i (\K + \p - \p') \cdot \bar{\x}_r}
	e^{-i (\q - \q' - \K) \cdot \bar{\x}_s'} e^{-\sigma_0^2 |\K + \p - \p'|^2} e^{-\sigma_0'^2 |\K + \q' - \q|^2}
	\notag \\
	& \times  \bar{u}_{s'}(\q') (I \otimes \sigma^3) u_s(\q) \bar{u}_{r'}(\p') (I \otimes \sigma^3) u_r(\p)
	\\
	& \times \! \Big[ \big\langle a_{r'}^\dag(\p') a_r(\p) a_i^\dag(\k) a_j(\k) \big\rangle
	\big\langle b_{s'}^\dag(\q') b_s(\q) b_k^\dag(\k) b_l(\k) \big\rangle 
	- \big\langle a_i^\dag(\k) a_j(\k) a_{r'}^\dag(\p') a_r(\p) \big\rangle 
	\big\langle b_k^\dag(\k) b_l(\k) b_{s'}^\dag(\q') b_s(\q) \big\rangle \Big] \notag.
\end{align}
The expectation values in the above equation are found using the general expression $\langle c_{r'}^\dag(\p') c_r(\p) c_{s'}^\dag(\q') c_{s}(\q) \rangle \approx (2\pi)^6 \delta_{\sigma_0}(\p-\q')\delta^3_{\sigma_0}(\q-\p')\delta_{rs'}\rho_{r's}(\q)$, see e.g. Refs.~\cite{Kosowsky:1994cy,Bavarsad:2009hm,Bartolo:2018igk,Bartolo:2019eac,Hoseinpour:2020hic,Zarei:2021dpb}.
One then integrates over the momenta $\p$, $\p'$, $\q$ and $\q'$ to find
\begin{align}
	\dot{\rho}_{IJ} &= \!-\frac{i \kappa^2}{4}  \hspace{-1.5mm}\sum_{rr'ss'}\hspace{-1mm}
	\int \! d\K\frac{q^0 p^0}{|\K|^2} 
	e^{-i \K \cdot (\bar{\x}_r - \bar{\x}_s')} e^{-(\sigma_0^2 + \sigma_0'^2) |\K|^2}
	\bar{u}_{s'}(\k) (I \otimes \sigma^3) u_s(\k)  
	\bar{u}_{r'}(\k) (I \otimes \sigma^3) u_r(\k) \notag \\
	&\hspace{5cm} \times  
	\big[ \delta_{ri} \delta_{sk} \rho^A_{r'j}(\k) \rho^B_{s'l}(\k) - \delta_{jr'} \delta_{ls'} \rho^A_{ir}(\k) \rho^B_{ks}(\k) \big] \,.
\end{align}
Here, the interaction induces mixing between the components of $\rho^A$ and $\rho^B$. As we will demonstrate later, this mixing gives rise to quantum entanglement.
Given the finite width of the Gaussian function in the above equation the integration over $\K$ needs to be done carefully.
In the limit of $\sigma_0 |\mathbf{K}| \ll 1$, which implies that the particles are well-localized, the Fourier transform over $\mathbf{K}$ results in
\begin{equation} 
	\dot{\rho}_{IJ} =- \frac{i \kappa^2}{16\pi} m_1m_2F(i,j,k,l)~,
\end{equation} 
where we have replaced $p^0 = m_1$, $q^0 = m_2$, and introduced the $F$ function with folded indices. The index mapping follows the tensor product rule, i.e., $I=2(i-1)+k$ and $J=2(j-1)+l$. The function is given by the following
\begin{equation} 
	F(i,j,k,l)=
	\sum_{rr',ss'}  \frac{1}{R_{sr}}\bar{u}_{s'}(\k) (I \otimes \sigma^3) u_s(\k)
	\bar{u}_{r'}(\k) (I \otimes \sigma^3) u_r(\k)  [ \delta_{ri} \,\delta_{sk} \,  \rho^A_{r'j}(\k) \rho^B_{s'l}(\k)
	- \delta_{jr'} \, \delta_{ls'} \, \rho^A_{ir}(\k)\rho^B_{ks}(\k) ] \, .  \label{ttdynamics}
\end{equation}
where, $R_{sr} = |\bar{\mathbf{x}}_r - \bar{\mathbf{x}}'_s|$ is the distance between the corresponding spatial states, which in turn are dictated by the spin states. The indices in the right- and left-hand sides of Eq.~\eqref{ttdynamics} are connected through the tensor product rule.
Note that considering the full form of the Gaussian distribution function for the particles one has corrections, see appendix \eqref{appendix:D} for the details.
The $\rho_{IJ}$ matrix elements are then obtained by solving the following joint differential equations ($\hbar=1$)
\begin{eqnarray}
	\dot\rho_{11}&=&0 \\
	\dot\rho_{12}&=&i \frac{\Delta x}{d(\Delta x+d)}\,\rho^A_{11}\rho^B_{12}\equiv \frac{\Delta x}{d(\Delta x+d)}\,\rho_{12} \\
	\dot\rho_{13}&=& -i\frac{\Delta x}{d(d-\Delta x)}\,\rho^A_{12}\rho^B_{11}\equiv -i\frac{\Delta x}{d(d-\Delta x)}\,\rho_{13} \\
	\dot\rho_{14}&=&0 \\
	\dot\rho_{22}&=&0 \\
	\dot\rho_{23} &=&-i\frac{2\Delta x}{d^2-\Delta x^2}\,\rho^A_{12}\rho^B_{21}\equiv-i\frac{2\Delta x}{d^2-\Delta x^2}\,\rho_{23}  \\
	\dot\rho_{24}&=&-i \frac{\Delta x}{d(\Delta x+d)}\,\rho^A_{12}\rho^B_{22}\equiv -i \frac{\Delta x}{d(\Delta x+d)}\,\rho_{24} \\
	\dot\rho_{33}&=&0 \\
	\dot\rho_{34}&=&i\frac{\Delta x}{d(d-\Delta x)}\,\rho^A_{22}\rho^B_{12}\equiv i\frac{\Delta x}{d(d-\Delta x)}\,\rho_{34} \\
	\dot\rho_{44}&=&0~.
\end{eqnarray}

%Now, summing over spin indices yields the following differential equation ??
%\begin{eqnarray} 
%	\dot{\rho}_{IJ} &=&?? -\frac{i}{2}V' \kappa^2 m_1m_2 \sum_{rr',ss'}  \frac{1}{R_{sr}}\bar{u}_{s'}(\k) (I \otimes \sigma^3) u_s(\k)
%	\notag \\
%	&& \!\!\! \times
%	\bar{u}_{r'}(\k) (I \otimes \sigma^3) u_r(\k)  \left[ \delta_{ri} \,\delta_{sk} \,  \rho_{r'j}(\k) \otimes\rho_{s'l}(\k) \notag \right. \\
%	&& \quad - \left. \delta_{jr'} \, \delta_{ls'} \, \rho_{ir}(\k)\otimes \rho_{ks}(\k) \right] \, ,
%\end{eqnarray}
Therefore,the full solution is given by the following matrix
\begin{eqnarray} 
	\rho_{[IJ]}(\tau) = \frac{1}{4}
	\begin{pmatrix}
		1 & e^{-i\Delta \phi_{LR}} & e^{-i\Delta \phi_{RL}} & 1 \\
		e^{i\Delta \phi_{LR}} & 1 & e^{-i(\Delta \phi_{RL} - \Delta \phi_{LR})} & e^{i\Delta \phi_{LR}}  \\
		e^{i\Delta \phi_{RL}} & e^{i(\Delta \phi_{RL} - \Delta \phi_{LR})} & 1 & e^{i\Delta \phi_{RL}} \\
		1&e^{-i\Delta \phi_{LR}}& e^{-i\Delta \phi_{RL}} & 1
	\end{pmatrix}
	~.
\end{eqnarray}

%where $\Delta \phi_{RL}$ and $\Delta \phi_{LR}$ are given in appendix \eqref{appendix:spin}.

\section{Fourier transform and integration over transfer momentum \( \mathbf{K} \)} \label{appendix:D}
In this appendix, we explicitly evaluate the integration over the momentum transfer \( \mathbf{K} \) and perform its Fourier transform. The interaction Hamiltonian derived in the main text involves an integral over the graviton propagator \( D^{\mu\nu\mu'\nu'}(\mathbf{K}) \), which depends on the exchanged momentum between the two interacting particles.
In the main text, we have considered the limit \( |\K|\sigma_0 \ll 1 \), which corresponds to the case where the particles are well-localized in space, meaning their wave packets have finite spatial widths. Under this assumption, the integral takes the form
\begin{equation}
	I(\mathbf{d}) =  \int \frac{d^3K}{(2\pi)^3} \frac{e^{i \mathbf{K} \cdot \mathbf{d}}}{|\mathbf{K}|^2} e^{-\sigma_0^2 |\mathbf{K}|^2/2}~.
\end{equation}
The Gaussian factor \( e^{-\sigma_0^2 |\mathbf{K}|^2/2} \) effectively introduces an ultraviolet (UV) cutoff, regularizing short-distance divergences. Evaluating this integral using spherical coordinates, we obtain
\begin{equation}
	I(\mathbf{d}) = \frac{1}{ d} \text{erf} \left( \frac{d}{\sqrt{2} \sigma_0} \right)~,
\end{equation}
demonstrating that the interaction strength is modified by the error function
For distances much larger than the localization width (\( d \gg \sigma_0 \)), the error function asymptotes to unity, and we recover the familiar Coulomb-like behavior
\begin{equation}
	I(\mathbf{d}) \approx \frac{1}{4\pi d}, \quad \text{for } d \gg \sigma_0~.
\end{equation}
In contrast, for distances much smaller than the localization width (\( d \ll \sigma_0 \)), the error function can be expanded as \( \text{erf}(x) \approx 2x/\sqrt{\pi} \), leading to a suppression of the interaction
\begin{equation}
	I(\mathbf{d}) \approx \frac{d}{\sqrt{8\pi} \sigma_0^2}, \quad \text{for } d \ll \sigma_0~.
\end{equation}
This suppression reflects the fact that the interaction is effectively smeared over the finite spatial extent of the wave packet.

\section{ Model II} \label{appendix:E}

An alternative way to retain the graviton-mediated qubit-qubit interaction is to introduce an external electromagnetic field $A^\mu$ that enables the coupling of gravity to the qubit system.
This can be achieved by modifying the qubit part to $A^\mu\bar{\psi}\gamma^\nu\psi$.
To simplify this interaction in the non-relativistic limit, we employ the Gordon decomposition, yielding the approximation
\begin{equation}
	\bar{\psi}(x)\gamma_{\nu}\psi(x) \approx \frac{1}{2m_i}\partial^\mu\!\left(\bar{\psi}(x)\sigma_{\mu\nu}\psi(x)\right)~, \label{Gordon}
\end{equation}
where $m_i$ is the mass of the fermions, and the spin tensor is defined as $\sigma_{\mu\nu} = i[\gamma_\mu,\gamma_\nu]/2$. This formulation highlights the role of intrinsic spin in the gravitational interaction.
After some rearrangements we arrive at the following Hamiltonian for Model II
\begin{eqnarray}
	\hat{H}_{\textrm{int}}^{\rm (II)}&\approx&\frac{\kappa^2}{16m_1m_2}\int d\mathbf{x}\, d\mathbf{x}'\,  D^{\mu\nu\mu'\nu'}(\mathbf{x}-\mathbf{x}')
	\partial^{\lambda'} A_{\mu'}\bar{\psi}^-(\x')\sigma_{\nu'\lambda' }  \psi^+(\x')
	\partial^\lambda A_{\mu}\bar{\psi}^-(\x)\sigma_{\nu\lambda }  \psi^+(\x)
	\nonumber\\ & & \!\!\!\!\times\,
	VV'\delta_{\sigma_0}^3(\mathbf{x}-\bar{\mathbf{x}}_r)\delta_{\sigma'_0}^3(\mathbf{x}'-\bar{\mathbf{x}}'_s)~. \label{Hmodel2}
\end{eqnarray}
The forward scattering results for the dynamical Model II are derived analogously, with the main distinction that in the coupling term  $ m_1 m_2 $ is replaced by $ \tfrac{1}{4}\omega_0^{(1)}\omega_0^{(2)}$, where $\omega_0^{(i)} {=} B_3/m_i$ is the Larmor frequency with $B_i{=-}\epsilon_{ijk}\partial^j A^k$ the external magnetic field.
The induced phase due to the graviton propagator in the presenice of background magnetic field in SI units is given by
\begin{equation}
	\phi_{B} \approx \frac{Gq^2B_3^2\tau \hbar}{m^2dc^4}~. \label{E3}
\end{equation}
To compare the coupling strength of model II with model I, we assume particles in both medels having the same mass \( m \) and particle in model II is subjected to a uniform background magnetic field of
\( B_3=1 \) T.
It is straightforward to confirm that the two couplings become comparable when in natural units one has
% \begin{equation}
	$m = \sqrt{B_3}$
% \end{equation}
which corresponds with the mass $m\sim 10^{-35}\,\text{Kg}$ is SI units. 
Consequently, for the magnetic fields around  \( B_3=1 \) T, Model II is definitely subdominant.  whereas for smaller masses, Model II provides a stronger coupling, and thus, a larger phase. The contribution from magnetic feild becomes dominant for magnetic fields larger than $B_3\sim 10^7 $T availabel in neutron stars.

\section{ Entanglement induced by photon propagator} \label{appendix:F}

\begin{figure}
\includegraphics[width=0.4\textwidth]{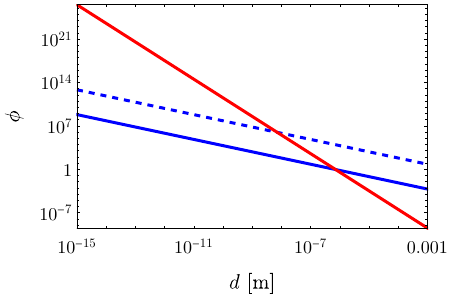}
 \caption{Comparing the phase accumulated within $\tau=1$ s through the gravitational (blue) and spin-spin (red) interactions as a function of the distance. For the gravity we consider particle mass of $m=10^{-15}$ kg (solid line) and $m=10^{-13}$ kg (dashed line), while for the spin-spin interaction electrons that have the highest gyromagnetic ratio in the nature are considered.}
 \label{fig:couplings}
\end{figure}

 The interaction between two qubits represented as a localized wavepacket of a spinor field $\psi(x)$ mediated by photon propagator is described by 
\begin{equation}
	\hat H_{\mathrm{int}}
	= VV' \int d\tau\, d\tau'\, d^{4}x\, d^{4}x'\,
	A_{\mu}(x)\, A_{\nu}(x')\,
	\bar{\psi}^{-}(x)\gamma^{\mu}\psi^{+}(x)\,
	\delta_{\sigma_0}^{4}\!\left(x-\bar x(\tau)\right)\,
	\bar{\psi}^{-}(x')\gamma^{\nu}\psi^{+}(x')\,
	\delta_{\sigma_0'}^{4}\!\left(x'-\bar x'(\tau')\right)~,
	\label{H_int_model}
\end{equation}
where the Wick contraction of the photon fields yields the Feynman propagator
\begin{equation}
	P_F^{\mu\nu}(x-y) = \langle 0 | T \{ A^\mu(x) A^\nu(y) \} | 0 \rangle
	= -i \eta^{\mu\nu} P_F(x-y)~,
\end{equation}
where its Fourier transform  is
\begin{equation}
	P_F^{\mu\nu}(x) = \int \frac{d^4 K}{(2\pi)^4} \, e^{-i K \cdot x} \, P_F^{\mu\nu}(K)
	= \int \frac{d^4 K}{(2\pi)^4} \, e^{-i K \cdot x} \, \frac{-i \, \eta^{\mu\nu}}{K^2 + i\epsilon}~.
\end{equation}
To simplify, define the localized currents
\begin{equation}
	\mathcal{J}^{\mu}(\tau)
	= \int d^{4}x\;
	\delta_{\sigma_0}^{4}(x-\bar x(\tau))\,
	\bar\psi^{-}(x)\gamma^\mu\psi^{+}(x)~,
\end{equation}
\begin{equation}
	\mathcal{J}'^{\nu}(\tau')
	= \int d^{4}x'\;
	\delta_{\sigma_0'}^{4}(x'-\bar x'(\tau'))\,
	\bar\psi^{-}(x')\gamma^\nu\psi^{+}(x')~.
\end{equation}
Then the Hamiltonian reduces to a worldline interaction
\begin{equation}
	\hat H_{\mathrm{int}}
	= VV'\!\int d\tau\, d\tau'\;
	\mathcal{J}^\mu(\tau)\,
	iP_{F\,\mu\nu}\!\left( \bar x(\tau)-\bar x'(\tau') \right)\,
	\mathcal{J}'^{\nu}(\tau')~.
	\label{Heff_worldline}
\end{equation}
Puting the effective Hamitonian into the forward scattering of the QBE \eqref{QBE1} in Fourier space one can show that the induced phase due to the Columb interaction is zero 
\begin{equation}
		\phi_{\text{C}} =0~.
\end{equation}
However,  the induce phase due to the spin-spin interaction is also agive as 
\begin{equation}
	\phi_{\text{SS}} \approx  
	\frac{q^2 \hbar^2\tau}{ m^2 d^3c} ~.
\end{equation}
In Fig.~\ref{fig:couplings} the spin-spin interaction is compared to the gravitational interaction.

%%%%%%%%%%%%%%%%%%%%%%%%%%%%%%%%%%%%%%%%%%

\end{document}